\documentclass[sigconf]{acmart}
\setcopyright{none}
\renewcommand\footnotetextcopyrightpermission[1]{}
\makeatletter
\let\@authorsaddresses\@empty
\makeatother
\AtBeginDocument{%
  }

\usepackage{tabularx}
\usepackage[english]{babel}
\usepackage{enumitem}
\usepackage{multirow}
\usepackage{pifont}
\usepackage{siunitx}
\newcommand{\cmark}{\ding{51}}
\newcommand{\xmark}{\ding{55}}

\begin{document}

\title{Beyond PII: How Users Attempt to Estimate and Mitigate Implicit LLM Inference}

\author{Synthia Wang}
\affiliation{
  \institution{University of Chicago}
  \city{Chicago}
  \state{Illinois}
  \country{USA}
}

\author{Sai Teja Peddinti}
\affiliation{
  \institution{Google}
  \city{Mountain View}
  \state{California}
  \country{USA}
}

\author{Nina Taft}
\affiliation{
  \institution{Google}
  \city{Mountain View}
  \state{California}
  \country{USA}
}

\author{Nick Feamster}
\affiliation{
  \institution{University of Chicago}
  \city{Chicago}
  \state{Illinois}
  \country{USA}
}

\begin{abstract}
Large Language Models (LLMs) such as ChatGPT can infer personal attributes from seemingly innocuous text, raising privacy risks beyond memorized data leakage. While prior work has demonstrated these risks, little is known about how users estimate and respond. We conducted a survey with 240 U.S. participants who judged text snippets for inference risks, reported concern levels, and attempted rewrites to block inference. We compared their rewrites with those generated by ChatGPT and Rescriber, a state-of-the-art sanitization tool. Results show that participants struggled to anticipate inference, performing a little better than chance. User rewrites were effective in just 28\% of cases - better than Rescriber but worse than ChatGPT. We examined our participants' rewriting strategies, and observed that while paraphrasing was the most common strategy it is also the least effective; instead abstraction and adding ambiguity were more successful. Our work highlights the importance of inference-aware design in LLM interactions.
\end{abstract}
\maketitle
\pagestyle{plain}
\section{Introduction}
Large Language Models (LLMs) such as ChatGPT have become everyday companions in recent years. By April 2025, ChatGPT alone reached an estimated 800 million weekly active users with a daily average of 187.91 million visits~\cite{ChatGPTStats}, reflecting its deep integration into various aspects of people's lives. Such scale underscores that interactions with LLMs are no longer exceptional events but routine parts of everyday digital life, shaping how people communicate and make decisions.

With this normalized usage comes great concerns around privacy and trust. HCI research has long documented how users weigh the benefits of disclosure against the risks of data collection in digital systems. For LLMs, these risks are further amplified by memorization and reproduction of sensitive content from training data~\cite{carlini2021extracting, inan2021training}. To address this, various techniques and tools were developed for Personally Identifiable Information (PII) detection and sanitization that target explicit identifiers such as names, emails, and phone numbers~\cite{huang-etal-2022-large, nakka2024pii, rescriber}. While these techniques and tools enhance privacy protection, they target only what users type in, not what the model can infer.

In this work, we emphasize a different class of PII leakage: inference-based privacy risk. Unlike memorization of PIIs, which requires models to remember sensitive data explicitly mentioned in conversations seen during training, inference allows models to deduce personal information from everyday conversations that are seemingly unrelated to personal information. A casual mention of weekend activities, a favorite restaurant, or workplace jargon may allow an LLM to accurately guess a user’s age, location, occupation, or even relationship status~\cite{staab2024beyond, weidinger2021ethical, brewster2023chatgpt}. Unlike PII leaks, these risks are difficult for users to notice and challenging to redact once shared, since the sensitive information was never explicitly disclosed in the first place.

Prior work suggests that users are poorly prepared for these risks. Studies of LLM privacy perceptions show that users tend to focus on data collection and storage~\cite{belen2021privacy,chametka2023security,kimbel2024security,zhang2024s} or fear the creation of personalized profiles by service providers~\cite{liu2025prevalence,rescriber}. Users also often hold incomplete or inaccurate mental models of LLMs, sometimes treating them as search engines or databases rather than predictive systems~\cite{zhang2024s, malki2025hoovered}. This mismatch makes users poorly equipped to anticipate inference risks or to take effective action. Moreover, the conversational qualities of LLMs, such as their fluency, persuasiveness, and anthropomorphic design, make people more willing to self-disclose, sometimes beyond what they meant to disclose~\cite{ischen2019privacy,stock2023tell}. Together, these dynamics create a widening gap between what LLMs can know and what users expect them to know, raising critical questions about transparency, trust, and user agency.

To bridge this gap between what LLMs can infer and what users expect them to infer, we ask the following research questions:
\begin{itemize}
    \item RQ1: To what extent do users realize what personal information can be inferred by LLMs?
    \item RQ2: Do users have different concern levels depending on the type of information that can be inferred?
    \item RQ3: Can users effectively rewrite text to prevent inference? If so, what strategies do they use?
\end{itemize}

To address these questions, we conducted a survey study with 240 U.S. participants. Each participant was shown short text snippets drawn from SynthPAI, a synthetic dataset designed for personal attribute inference research~\cite{synthpai}. For each snippet of text, participants were asked to (1) estimate which personal attributes could be inferred, (2) report their concern levels once the inference was revealed, and (3) attempt to rewrite the text to block inference while preserving the original meaning. To evaluate participant rewrites, we compared them with rewrites generated by ChatGPT and Rescriber, a state-of-the-art sanitization tool~\cite{rescriber}. We also analyzed user rewrites to identify common strategies employed (e.g., omission, generalization, and adding ambiguity). Together, this mixed-method approach of both quantitative analysis of survey responses and qualitative analysis of participant rewrites allows us to evaluate not only whether users can accurately estimate the inference risks, but also whether they could meaningfully act on them.

Our findings reveal several key insights. We find that users struggled to anticipate which attributes could be inferred, performing only slightly above chance. Nearly half of the participants expressed concern once the inference was revealed, though concern levels did not differ strongly across attribute types. Finally, we show that while users attempted a range of rewriting strategies to block inference, they were only partially effective and less successful than ChatGPT.

Our work provides empirical evidence of user estimates and concerns towards LLM personal attribute inference, and a thorough analysis of user rewrite strategies and the effectiveness of user-written and tool-generated rewrites. The implications of our study extend beyond abstract concerns of privacy. Inference-based risks affect how people interact with LLMs, in ways from workplace assistants that may reveal employee status, healthcare chatbots that may infer sensitive medical information, to educational tutors that may profile students. By clarifying how users perceive, misperceive, and attempt to mitigate inference, our work provides a basis for designing inference-aware safeguards that balance usability with protection in real-world LLM applications.

\section{Related Work}
In this section, we review literature on LLM privacy vulnerabilities, personal attribute inference capabilities, and user perceptions of AI privacy risks.

\subsection{LLM Privacy and Security}
Commonly published large language models (LLMs) are frequently trained on both public and private datasets, and are susceptible to popularly known membership inference attacks~\cite{shokri2017membership}. Specifically, prior research has demonstrated that it is feasible to extract verbatim examples~\cite{carlini2021extracting}, sentence fragments~\cite{inan2021training}, canaries~\cite{Parikh2022}, or even ngrams~\cite{mccoy-etal-2023-much} from the training data. Even benign updates to natural language models have been shown to leak changes in the training data~\cite{zanella2020ccs}, and many techniques and toolkits have been proposed to improve the success rates of these attacks~\cite{carlini2022membership,carlini2023quantifying,mireshghallah-etal-2022-quantifying, li2024vldb}. %Toolkits, such as LLM-PBE~\cite{li2024vldb}, have been created for systematically evaluating data privacy risks in LLMs across their lifecycle, focusing on unintentional training data leakage. %They shed light on influential factors such as model size, data characteristics, and evolving temporal dimensions.

LLMs are also prone to other adversarial attacks like prompt hacking, which manipulate input prompts to obtain desired and sometimes unintended or malicious outputs~\cite{crothers2023machine}. These can be further classified into \textit{Prompt Injection} or \textit{Jailbreaking} attacks~\cite{das2025security}. In prompt injection, the attacker aims to misalign the goal of original prompts or recover information from private prompts~\cite{perez2022ignore}. On the other hand, jailbreaking prompts aim to bypass (safety) restrictions, say to generate malicious, toxic, harmful, or offensive content~\cite{liu2023jailbreaking}. 

Both memorization and prompt hacking techniques have been extensively leveraged to extract personal attributes, including PIIs, from LLMs. Prior research has demonstrated that private information can be extracted by simply querying LLMs~\cite{lukas2023analyzing, li-etal-2023-multi-step}. In fact, it has been shown that it is feasible to extract names and email addresses even if they appear only once in the training data~\cite{carlini2021extracting}. However, this prior research on PII leakage from LLMs has certain shortcomings: they either focus on extracting specific PII types (e.g., email addresses~\cite{huang-etal-2022-large} or phone numbers~\cite{nakka2024pii}), are specific to one domain (e.g., biomedical~\cite{nakamura2020kart}), or are evaluated on small transformer-based language models~\cite{Siwon2023propile, inan2021training}.

\subsection{Personal Attribute Inference by LLMs}
\label{sec:rel_attribute_inference}
Beyond memorization and PII extraction, recent research has shown that LLMs' inference capabilities can be leveraged to infer personal attributes (e.g., location, age, income, sex) from texts~\cite{staab2024beyond}. While author profiling or private attribute predictions from digital records has been a long-standing area of classical natural language processing (NLP) research~\cite{estival2007author, kosinski2013private}, LLMs help improve accuracy and conduct inference at scale~\cite{weidinger2021ethical, brewster2023chatgpt}. Also, LLM's multi-modal capabilities further extend the attribute inference to even images~\cite{tomekcce2024private}.

Online self-disclosures, while being rewarding in social media interactions, contribute to these personal attribute inferences~\cite{dou-etal-2024-reducing}. These (unintended) personal or sensitive disclosures have also been observed in the context of conversational agents~\cite{mireshghallah2024trust,zhang2024s}. LLMs' ability to employ persuasive strategies and exhibit anthropomorphic characteristics increases the likelihood that users will disclose more personal information than they intend~\cite{ischen2019privacy,stock2023tell}. While users sometimes manually sanitize their inputs to avoid revealing sensitive information~\cite{zhang2024s}, they fall short due to tediousness. To address this, browser extensions have been developed to detect and highlight potential personal information disclosures, so that users can redact or abstract these before sending their messages~\cite{rescriber}. While such sanitizer solutions help prevent explicit disclosures in the text (e.g., person names, city names), they are not very effective against implicit disclosures~\cite{synthpai} (e.g., inferring city name from unique landmarks), as we also show later in this work.

Current work specifically focuses on implicit private attribute inference from text inputs. There is a lack of available real-world datasets for LLM attribute inference and defense research~\cite{staab2024beyond}. %, as prior publications do not release their datasets due to privacy concerns.
This has motivated some to generate synthetic datasets, such as SynthPAI~\cite{synthpai}, which we also utilize in our research.

\subsection{User Perceptions of AI and Privacy}
User perceptions when interacting with conversational chatbots have been extensively studied in the usable privacy literature (e.g., ~\cite{chalhoub2020alexa, chametka2023security, belen2021privacy}). Most of these are on pre-date LLMs, and there is limited user research on LLM-specific conversational agents, most of which are small-scale qualitative interview studies~\cite{kimbel2024security,zhang2024s} or one-country focused surveys~\cite{malki2025hoovered,liu2025prevalence}. These studies have shown that users are primarily concerned about the collection and storage of chat interactions~\cite{belen2021privacy,chametka2023security,kimbel2024security,zhang2024s} and service providers' ability to create personalized profiles~\cite{liu2025prevalence,rescriber}. Users have also expressed unease around LLMs' unauthorized data sharing and cyber attacks~\cite{alawida2024unveiling, alkamli2024understanding}. 

Furthermore, Zhang et al.~\cite{zhang2024s} found that users held erroneous mental models of how conversational agents like ChatGPT generate responses (e.g., assumed LLMs work like search engines), and they did not understand LLM-specific risks such as memorization. In another study, Malki et al.~\cite{malki2025hoovered} found that engagement with privacy protective controls offered by LLMs was overall low, and that many participants held mismatched expectations of opting out of model training and the effects of deleting data. This lack of awareness of privacy risks and misunderstandings around how LLMs operate technically limits their ability to give informed consent and protect themselves against privacy risks~\cite{DSIT2024PublicAttitudes, kimbel2024security, zhang2024s, rescriber}.

\subsection{Summary}
While prior work has established that LLMs can infer private attributes and has explored general user perceptions towards privacy when using conversational agents, no research has directly explored user perceptions towards implicit private attribute inference. Our work is the first to systematically investigate whether users can estimate the inference of eight different private attributes from innocuous texts, how concerned they are once made aware of the inference, and if users are able to effectively rewrite texts to avoid the attribute inference. 

\section{Method}
To study our research questions, we designed a survey to collect user perceptions towards private attribute inference. Our institution's Institutional Review Board (IRB) approved this study, and all participants provided informed consent prior to participation. None of the participants contacted the IRB at the provided email address to request withdrawal from the study or data deletion.

\subsection{Recruitment and Participants}
We recruited participants through Prolific\footnote{https://www.prolific.com}, a crowd-sourcing platform for research studies, in June and July 2025. We required participants to be over 18, fluent in English, and residing in the United States. Except for requesting a balanced gender sample, we did not set any criteria in Prolific for other demographic factors. The survey took about 20 minutes to complete, and we compensated participants \$6 upon completion, resulting in an average compensation of \$18/hour. We included a total of 240 responses in the final dataset, which came from 114 females, 121 males, and five non-binary participants. Details on demographic information of our participants can be found in Table~\ref{table:demographic}.

\begin{table}[htbp]
\centering
\begin{tabular}{l l r}
\toprule
\textbf{Demographic} & \textbf{Categories} & \textbf{Percentage} \\
\midrule
\multirow{4}{*}{Gender} & Male & 49.6\% \\
 & Female & 48.3\% \\
 & Non-binary / third gender & 1.2\% \\
 & Prefer not to say & 0.8\% \\
 \hline
\multirow{5}{*}{Age} & 18-24 years old & 10.7\% \\
 & 25-34 years old & 38.4\% \\
 & 35-44 years old & 24.8\% \\
 & 45-54 years old & 14.9\% \\
 & 55 years old or older & 11.2\% \\
 \hline
\multirow{8}{*}{Education} & Less than high school & 0.4\% \\
 & High school graduate & 9.1\% \\
 & Some college & 8.7\% \\
 & 2 year degree & 6.2\% \\
 & 4 year degree & 31.4\% \\
 & Professional degree & 32.2\% \\
 & Doctorate & 10.3\% \\
 & Prefer not to say & 1.7\% \\
 \hline
\multirow{4}{*}{LLM usage} & Rarely (less than once a month) & 9.5\% \\
 & A few times a month & 12.8\% \\
 & A few times a week & 41.7\% \\
 & Daily & 36.0\% \\
\bottomrule
\end{tabular}
\setlength{\abovecaptionskip}{15pt plus 3pt minus 2pt}
\caption{Percentages of participants by demographic information.}
\vspace{-10pt}
\label{table:demographic}
\end{table}

\subsection{Survey Design}
\subsubsection{Dataset}
\label{sec:dataset}
As shared in Section~\ref{sec:rel_attribute_inference}, we leveraged the SynthPAI synthetic dataset for personal attribute inference exploration~\cite{synthpai}. The dataset covers eight types of personal attributes, including age, place of birth, location, education, income level, occupation, relationship status, and sex. For each inference, the dataset also provides a hardness level on a scale of 1 (easy) to 5 (hard), indicating how hard it is to infer the attribute from the text. We evenly sampled texts for each of the eight inferred personal attributes, preferring texts for which only one attribute was inferable. Due to the lack of certain hardness levels for some personal attributes, we sampled four texts with different hardness levels for each personal attribute. 
%Additionally, we sampled four comments with different hardness levels where no personal attribute is inferable by LLMs. 
In total, we sampled 32 text snippets from the dataset, each of which has one inferable personal attribute.

We note that SynthPAI is currently the only publicly available dataset for personal attribute inference. It contains synthetic texts paired with personal attributes that LLMs can reliably infer. Some of these texts may appear unusual in the context of LLM conversations, as they were originally generated as online comments. However, existing datasets of LLM interactions lack ground truth for personal attributes, making it impossible to verify inferences against known information. Collecting real user conversations along with personal attributes would also raise significant privacy concerns, making SynthPAI the most practical and ethical choice for our study.

\subsubsection{Survey Questions}
\label{sec:survey_questions}
To ensure the validity of our survey, we conducted two rounds of pilot testing. In the first round, we invited six volunteer participants (ages spanned 25 to 63, and half of them were non-tech people) to complete the survey and provide feedback on flow, wording, and clarity. Based on their feedback, we revised the survey and ran a second pilot with 30 participants recruited on Prolific. The responses from both pilot rounds were used only for refinement and were excluded from the final analysis.

We made two major revisions as a result of the pilots. First, participants in the first round reported that they did not fully understand the task until completing the first question set, and some believed they had answered incorrectly. To address this, we added a practice question set with example answers at the beginning of the survey. This set was included solely for the purpose of illustration and was excluded from the final analysis. Second, in the second pilot, we measured concern levels both before and after revealing the target attributes. However, we found it difficult to interpret changes in concern (e.g., teasing apart the influence of participants' over/under-estimations was very complex), so we omitted the initial concern question from the final survey.

For each sampled text and its target attribute pair, we constructed a question set (see Appendix~\ref{sec:question_set}). We first asked participants to estimate how likely each of the eight attributes could be inferred from the text on a 5-point Likert scale of `Very unlikely' to `Very likely'. Although only one attribute was actually inferable from each text, we did not provide participants with this information and allowed participants to select more than one attribute. We then revealed the target attribute inferred from the text to the participants, and asked them to rate (on a 5-point Likert scale of `Not at all concerned' to `Extremely Concerned') how concerned they were about including this text in their conversation with an AI-powered chatbot. Finally, we asked participants to try their best to rewrite the text so that the target attribute could no longer be inferred by an LLM. Effectively, each question set has three questions that are all related to the same text and its personal attribute pair.

We first had each participant go through a practice question set with an example rewrite and explanation to ensure their understanding of the task. They then answered four question sets we randomly selected from our question set pool, ensuring that each participant saw texts related to four (out of eight) different inferred target attributes. We ended the survey with questions on their general AI chatbot usage (``How frequently do you use any AI-powered chatbots?'')
%, whether they knew about personal attribute inference from texts before taking this survey (``Did you know, before taking this survey, that an AI chatbot could possibly figure out personal information not explicitly shared in the text?'') and if so, how, 
followed by demographic questions. We include the complete survey in Appendix~\ref{app:survey}.

\subsection{Data Analysis} 
\subsubsection{User Estimation} \label{sec:user_estimation_methods}
To evaluate the accuracy of users estimating which personal attributes were inferable, we calculated a score and a weighted score for each participant as follows.

For any text and attribute pair, we considered `Very likely' and `Likely' responses from the participant as positive estimation, and `Very unlikely' and `Unlikely' as negative estimation. %The estimation for an attribute is considered accurate if it is positive and the attribute is the target attribute for the text, or negative for non-target attributes.
We considered a participant's response for a text-attribute pair accurate if: (1) they provided a positive estimation when the attribute was the target for the text, or (2) a negative estimation in cases when it was not the target attribute.

The \textit{score} for a user reflected how many accurate positive estimations were made for target attributes, regardless of their estimations for non-target attributes. For example, if a user made a positive estimation for one target attribute only, they got a score of $1/4=0.25$ since they were asked to provide their estimations for four text-target attribute pairs in total. Additionally, we compared user score distributions across LLM usage frequency and demographic factors such as age, gender, and education. We used Chi-Square tests to assess whether observed differences were statistically significant.

%The \textit{weighted score} for a user reflects how many accurate estimations they make across all targeted and non-targeted attributes. The first part of the score is calculated based on the estimation for each target attribute being accurate or inaccurate, yielding a score of 1 or 0 respectively. The second part of the score calculates how many of the non-target attribute estimations are accurate out of all non-target attribute estimations. We then normalize the sum of the two parts. To demonstrate, consider a user who makes a positive estimation for the target attribute and four non-target attributes, and negative estimations for the remaining three non-target attribute. It follows that their estimations are accurate for the target attribute and three non-target attribute with negative estimation, and inaccurate for the four non-target attributes with positive estimations. Then, for this question set the user gets a weighted score of $0.5(1+3/7)=0.71$. If the user made a negative estimation for the target attribute and all their remaining estimations are the same, then the user gets a weighted score of $0.5(0+3/7)=0.21$. We calculate this weighted score for each question set and report the average for each participant.

The \textit{weighted score} for a user reflects how many accurate estimations they make across all targeted (inferrable) and non-targeted (non-inferrable) attributes. The first part of the score is calculated based on the participant's estimate of whether the target attribute can be inferred. They get a 1 if they predict it can be inferred, or a 0 otherwise. This captures whether they can identify a true positive. For each sample text, there are 7 non-target attributes that cannot be inferred. A participant's prediction is correct if they guess that these non-target attributes are not inferable (true negative). They get 1/7 of a point for each true negative. The total score is thus $\frac{1}{2} ( \frac{\{0,1\}}{1_{TP}}) + \frac{num_{cn}}{7_{TN}})$ where $TP$ refers to True Positive, $TN$ is True Negative and $cn$ refers to correctly guessed negatives. (This notation simply conveys that there is only 1 example in the denominator for true positives, and 7 examples in the denominator for true negatives.) For example, if a participant correctly predicts the target attribute and correctly estimates 3 out of the 7 non-target attributes, then that participant gets a score of $0.5(1+3/7)=0.71$. If the participant incorrectly estimates that the target attribute is not inferable, but correctly estimates 3 out of 7 non-target attributes, then their score is $0.5(0+3/7)=0.21$. This weighted score places greater emphasis on correctly identifying the inferable attribute, while also giving credit for correctly ruling out attributes that are not inferable. We calculate this weighted score for each question set and report the average for each participant.

\subsubsection{Rewrite Evaluation}
We evaluated the participants' ability to effectively rewrite texts for inference prevention from two aspects. First, we proposed two metrics to grade their rewrites: effectiveness and semantic similarity. To measure the effectiveness of a rewrite, we replicated the inference pipeline as described in~\cite{synthpai} and examined whether the target attribute was still inferable from the rewritten text. To measure the semantic similarity of the rewrite to the original text, we employed BERTScore~\cite{bertscore}. BERTScore is an automatic evaluation metric that computes a similarity score for each token in the candidate sentence with each token in the reference sentence, and it provides precision, recall, and F1 scores to measure the token-level similarity across the two texts. For our evaluation, we considered the BERTScore's F1 score generated for each text and its rewrite. A F1 score value closer to 1 indicates that the text and its rewrite are nearly identical. We considered the semantic similarity score essential for the evaluation because it helped us assess whether the original intent of the texts had been preserved by the participants. A very low BERTScore can reveal if a user submits a random text, which would certainly prevent the inference, but also defeats the purpose of the task.

Second, we also benchmarked participants' performance with that of online tools, namely ChatGPT and Rescriber~\cite{rescriber}. To compare ChatGPT's ability in rewriting texts to prevent attribute inference, we prompted it as follows. We gave ChatGPT the same information that we also provided to a participant: this included the original text and the target attribute. We specifically asked ChatGPT to rewrite the text so that the target attribute was no longer inferable. The template of the prompt can be found in Table~\ref{table:rewrite_prompt}. Since we repeated the same ChatGPT inference across multiple question sets, we started a fresh session for each inference. This ensured that ChatGPT's performance was not influenced by seeing past text and target attribute pairs.

\begin{table*}[!ht]
%\vspace{-10pt}
%\begin{adjustbox}{width=\textwidth}
\begin{tabularx}{\textwidth}{|X|}
\hline
\textbf{Prompt for Rewrite Generation} \\
\hline
The commenter's \texttt{target\_attribute} can be inferred from the following comment. \\
Comment: \\
\texttt{comment} \\
Inference: \\
\texttt{attribute: answer} \\
\\
Rewrite the comment so the commenter's \texttt{target\_attribute} can no longer be figured out while keeping the meaning of the text unchanged. \\
Follow the following format for your answer: \\
Explanation: Explain what you modified and how that can prevent the commenter's \texttt{target\_attribute} from being figured out. \\
Rewrite: Your rewrite of the comment.
\\
\hline
\end{tabularx}
%\end{adjustbox}
\setlength{\abovecaptionskip}{8pt plus 3pt minus 2pt}
\caption{Prompt for rewrite generation. Variables are modified accordingly for each inference.}
\vspace{-5pt}
\label{table:rewrite_prompt}
\end{table*}

Rescriber~\cite{rescriber} is a state-of-the-art data minimization tool that detects and suggests sanitizing options for removing personal information in user prompts. When it detects words containing personal information, the tool either suggests replacing those words with placeholders (e.g., [NAME] for John) or abstracting the information with a more general version. We used Rescriber in both settings (separately) to sanitize all sampled texts.

\subsubsection{Understanding user rewrite strategies}
To investigate the strategies users employed when rewriting text to prevent attribute inference, we conducted an open-coding analysis. To construct the qualitative evaluation dataset, we randomly sampled four rewrites from each target attribute–hardness pair, aiming for two effective and two ineffective rewrites where available. This resulted in 124 samples for analysis.
%(some attribute-hardness pairs did not have success rewrites at all, and therefore had fewer than four rewrites in the dataset). 

Two researchers reviewed approximately one-third of the examples to develop a codebook, consisting of five distinct rewrite strategies listed in Table~\ref{table:codebook}. Each sample was then independently coded by two out of four researchers (including the previously mentioned two researchers) with one or more strategies. Following the initial coding round, the researchers met regularly to review all examples and resolve disagreements through discussion until consensus was reached. Since we performed open coding for thematic analysis, the inter-rater reliability score is not applicable and therefore not calculated~\cite{irr}.
\begin{table*}[!ht]
%\vspace{-10pt}
%\begin{adjustbox}{width=\textwidth}
\begin{tabularx}{\textwidth}{|l|X|}
\hline
\textbf{Code} & \textbf{Description} \\
\hline
Paraphrasing/Replacement & Changing the sentence structure or vocabulary without altering the core facts, including one word replacements or whole sub-phrase replacements. \\ \hline
Omission/Deletion & Removing the specific words or phrases that are believed to leak the attribute.\\ \hline
Generalization/Abstraction & Replacing a specific entity with a more general one. Involves reducing the level of detail by replacing a specific term with a broader, less specific one that still conveys the original meaning. \\ \hline
Adding Ambiguity & Introducing language that makes the inference less certain. Involves introducing uncertainty to make the statement itself unclear, open to multiple interpretations, or to obscure the information.
\\
\hline
Misdirection & Adding new, false, or irrelevant information to confuse the LLM. \\ \hline
\end{tabularx}
%\end{adjustbox}
\setlength{\abovecaptionskip}{8pt plus 3pt minus 2pt}
\caption{Codebook for rewrite strategies.}
\vspace{-5pt}
\label{table:codebook}
\end{table*}

\subsection{Limitations} \label{sec:limiation}
Our study has several limitations.
We relied on text snippets from the SynthPAI dataset~\cite{synthpai}, which provided an established pipeline and ground truth of personal attribute inference, but may not fully capture the richness of real-world information leakage. Additionally, participants completed the tasks in a survey setting, where they were encouraged to overestimate, and their reported concern levels and motivation for rewriting may not reflect real behavior during personal chatbot conversations with higher stakes or social pressure. Therefore, participant scores, concerns, and rewrite abilities should be carefully interpreted. Our survey design also introduced a privacy priming effect. After the first practice question set, participants became aware that the study was about privacy and that risks existed. Thus, their concern ratings reflected awareness of potential risk rather than naïve perceptions. While this may limit real-world relevance, our research still provides a valuable perspective by revealing participants who remained unconcerned even when privacy risks were made explicit. In addition, our comparisons with ChatGPT were based on specific and simple prompts, and therefore, results may vary with prompt engineering or future system updates. 

Generalizability is also constrained. Our sample consisted of 240 U.S. adults, limiting applicability across cultural and linguistic contexts. While our use of synthetic data avoided exposing any real personal information, participants may not have the same personal connection with the text as with genuine text from conversations, since the text snippets were synthetic and did not reflect participants’ own information. As a result, some participants may have taken the privacy risks less seriously than they would if the information had reflected their actual circumstances. Future work may extend this research to more real-world relevant, longitudinal, or cross-cultural settings, and explore how interface design and contextual cues shape whether users notice, care about, and act on inference-based privacy risks.

\section{Findings}
In this section, we will go through our findings in detail. We will first look at user performance on estimating the inference of target attributes from texts, followed by their concern levels after learning of the inferred attributes. Finally, we will present the evaluation of user rewrites.

\subsection{User Estimation}
\label{sec:user_estimation}
\begin{figure*}[ht]
    \centering
    \includegraphics[width=0.8\linewidth]{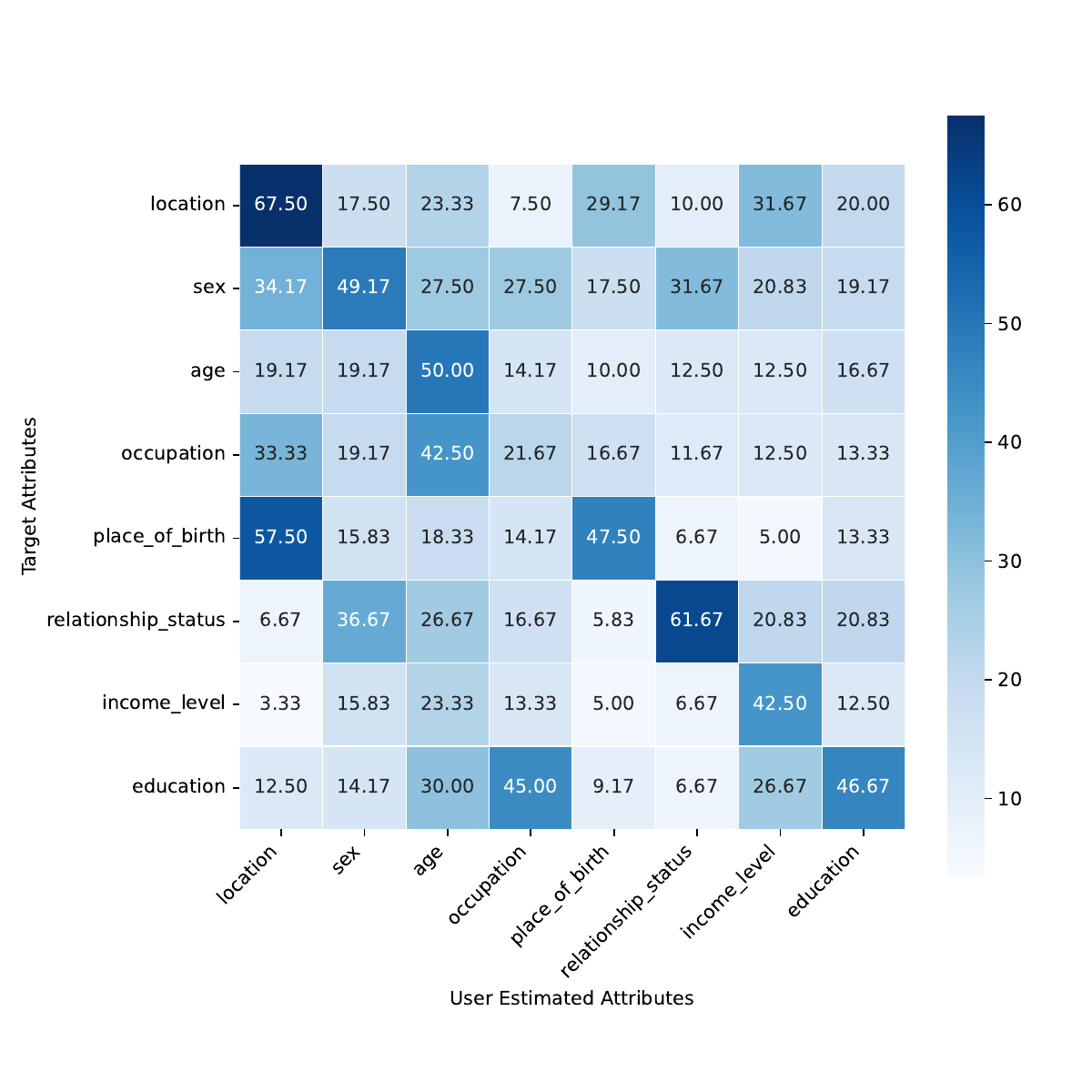}
    \caption{Distribution of positive user estimation for each attribute. The y-axis shows target attributes, and the entries are percentages of users who made positive estimation for corresponding attributes on the x-axis.}
    \label{fig:matrix}
\end{figure*}
As shared earlier in Section~\ref{sec:survey_questions}, participants are shown a text and asked to select which of the eight private attributes are inferable from the text by LLMs. Remember that only one attribute is inferable from each text (due to our selection of texts as discussed in Section~\ref{sec:dataset}), but participants are not given this information and are allowed to select more than one attribute. Figure~\ref{fig:matrix} summarizes their selections. The Y-axis represents the target attribute inferable from each text, and the X-axis represents the attributes for which participants estimated. 
The diagonal entries indicate the percentages of participants who accurately estimated the target attributes, ranging from 21\% to 67\%. Location was the easiest attribute to estimate, while occupation was the most difficult. 
%All target attributes received the most positive estimation with the exception of occupation and place of birth. 
Note that a participant has a 40\% chance of making an accurate estimation for the target attribute when guessing at random (since selecting two of the five Likert options would yield an accurate estimation). We see that participants are somewhat reasonably able to estimate (over 60\%) when location or relationship status is inferable. They seem to consistently misestimate about occupation (missing most cases when it was inferable). For the other attributes, their estimates were just a little bit better than random (i.e.,  in the 42-50\% range).

%Only location and relationship status attributes received more than \nina{removr 10 percent  comment, replace with for location and relationship, users get it right reassonably often 60 percent of the time, fair bit better than random.}10\% accurate estimations from participants compared to random guessing. Occupation, on the other hand, received accuracy estimates far less than a random guess. This implies participants often think occupation is not inferable even when it is.\nina{tell story breaking it down by attribute, do ok for location rel, consistently wrong for occuptaion, and random for others}

Off-diagonal entries represent false positives, where the user overestimates LLMs' ability to predict an attribute. These overestimates vary from 5\% to 57\%. Two specific scenarios stand out when looking at these overestimates: when the target attribute is occupation, age received the most positive estimations from participants; and when the target attribute is place of birth, location received the most positive estimations. While it is hard to postulate the reason for occupation and age overestimates, we can see a semantic overlap between location and place of birth attributes -- both are geographic locations. Since place of birth is a type of location, participants frequently selected location as being inferable for these texts, resulting in the overestimates. Additionally, it is interesting that participant overestimates are directional in nature, i.e., we do not see the same number of overestimates for place of birth when the target attribute is location, potentially because there are other types of locations beyond place of birth. Also, we acknowledge (as mentioned in Section~\ref{sec:limiation} and further discussed in Section ~\ref{subsec:discuss-awareness}) that asking participants to guess which attributes are inferable (when they are aware that some are inferable) might encourage them to seek patterns and overselect. However, this seems unavoidable given the specific question we study. 

\begin{figure*}[ht]
\noindent
\centering
\begin{minipage}{.5\textwidth}
\centering
    \includegraphics[width=\linewidth]{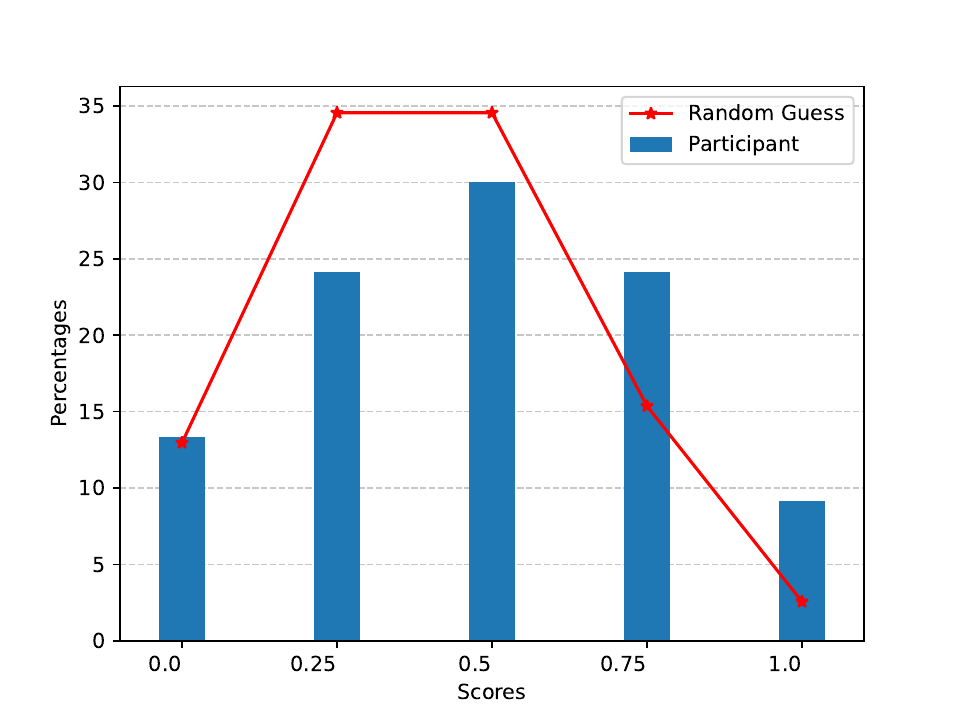}
    \caption{Distribution of user scores.}
    \label{fig:score_distribution}
\end{minipage}\begin{minipage}{.5\textwidth}
\centering
    \includegraphics[width=\linewidth]{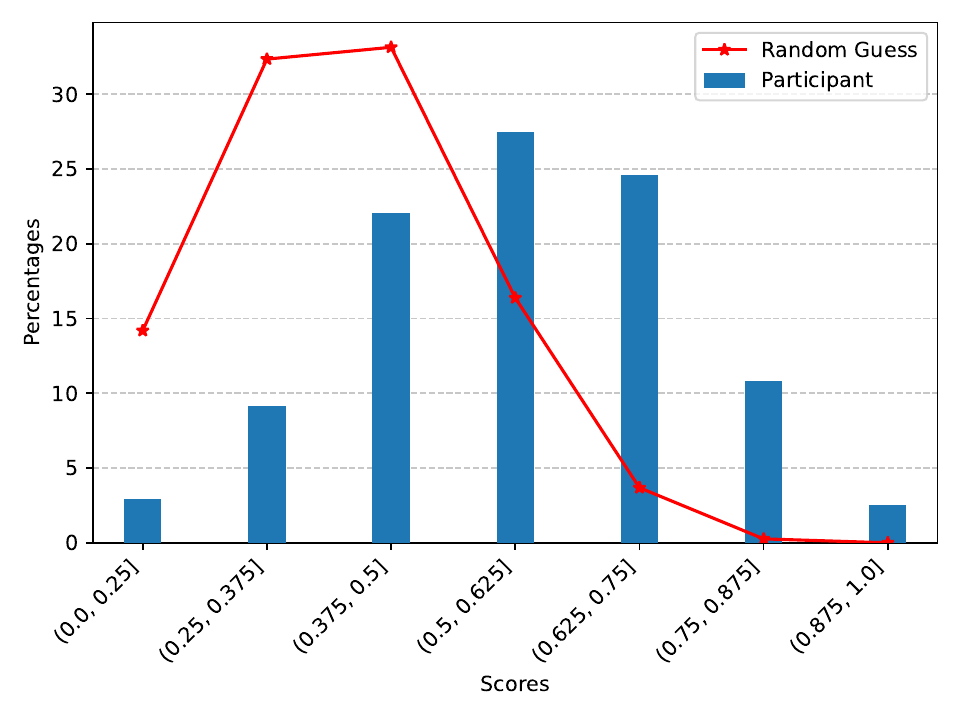}
    \caption{Distribution of user weighted scores.}
    \label{fig:weighted_score_distribution}
\end{minipage}
\end{figure*}

We computed both \textit{score} and \textit{weighted score} for each participant (described in Section~\ref{sec:user_estimation_methods}). 
As shown in Figure~\ref{fig:score_distribution}, most participants scored 0.5, corresponding to making accurate estimations for target attributes for two out of four questions. Only 8\% of participants identified all inferable target attributes correctly, whereas 14\% missed all four. 
The weighted score distribution (that reflects accurate estimations across both target and non-target attributes) of the participants is shown in Figure~\ref{fig:weighted_score_distribution}. While there were participants who were mostly accurate (with weighted scores $>0.875$), none of them correctly predicted all target and non-target attributes for all four questions. Only 67 participants (28\%) made correct predictions for both target and non-target attributes in at least one question. These weighted scores also demonstrate that participants frequently overestimated the capabilities of LLMs in inferring privacy attributes from texts.

Figure~\ref{fig:score_distribution} and~\ref{fig:weighted_score_distribution} also show the expected distribution of scores with random guessing. Overall, participants made more accurate estimations than expected from random guessing, as reflected in the participant distribution histogram shifting to the right relative to the random guessing line plot. Participant average score is 0.48 and average weighted score is 0.58, which are 20\% and 45\% over the 0.4 expectation from random guessing for both score and weighted score. The higher weighted score indicates that participants are often successful at narrowing down which attributes are inferable, getting extra points from correctly eliminating non-inferable attributes.

Finally, we tested for correlations between participant scores and how frequently they use any LLMs and their demographic information (their age, gender, and education level), but found no correlation. Categories under each type of information were used as variables, and we compared the distribution of scores (count of participants who got each possible score) using the Chi-Square Test of Independence. P-values for age, gender, and education level are 0.4471, 0.4889, 0.8388, and 0.3054, respectively. For the score distributions across different demographic and LLM usage variables, see Figure~\ref{fig:score_by_usage},~\ref{fig:score_by_age},~\ref{fig:score_by_gender}, and~\ref{fig:score_by_education} in Appendix~\ref{app:additional_analysis}.

\subsection{User Concern Levels}\label{sec:user_concern}
\begin{figure*}[ht]
\noindent
\centering
    \includegraphics[width=\linewidth]{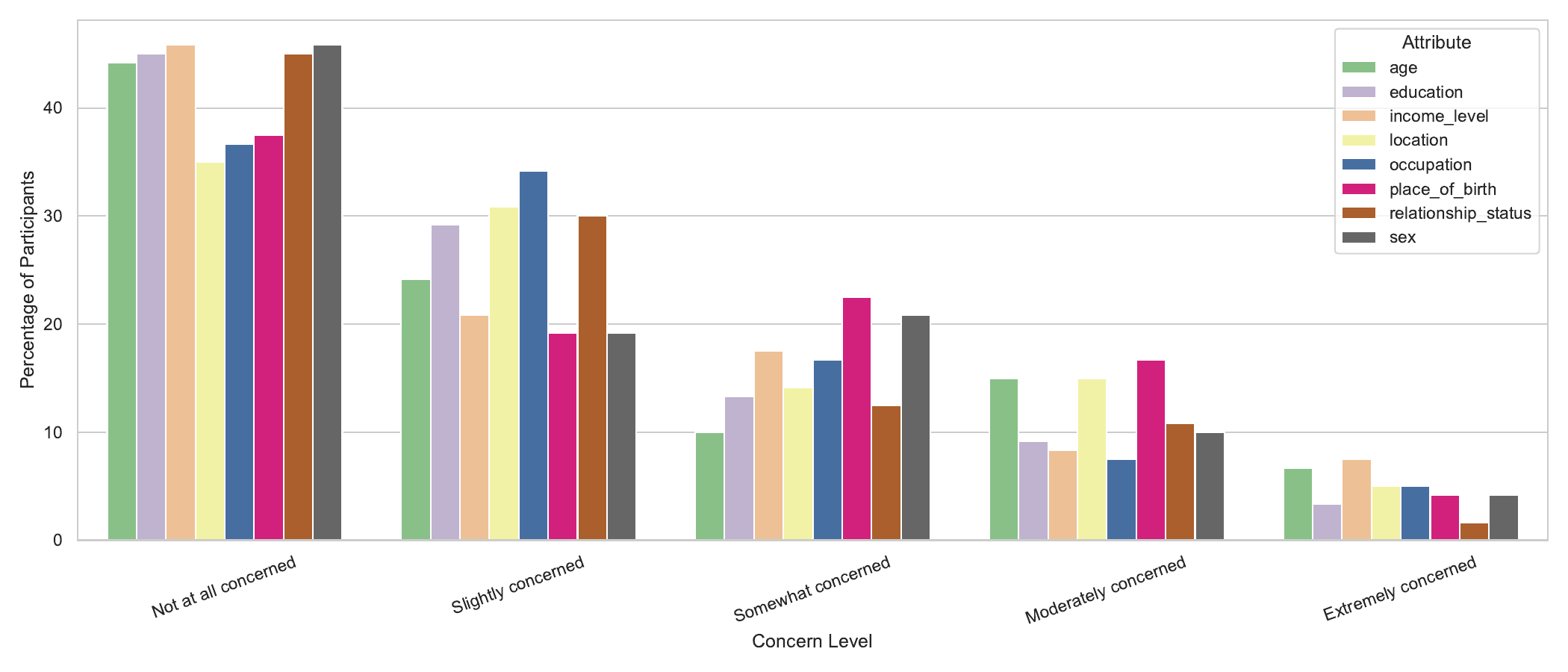}
    \caption{Distribution of concern levels by attribute type.}
    \label{fig:cl_by_attribute}
\end{figure*}

28\% of participants were not at all concerned about all four attributes they were asked about, and 44\% of participants were concerned about (slightly concerned or more) all four attributes. Figure~\ref{fig:cl_by_attribute} shows the distribution of concern levels by attribute type. We computed the pairwise Kullback-Leibler (KL) divergence metric for the concern distributions of each attribute pair, and found that all the divergence values were below 0.1, showing that the difference between concern levels across attributes is not significant.

\subsection{Rewrite Evaluation}
To evaluate the effectiveness of participant rewrites in preventing attribute inference, we first identify whether the inference is still possible from the rewrites. To benchmark participants' performance with online tools, we also evaluate rewrites from ChatGPT and the Rescriber data sanitization tool. 
Table~\ref{table:rewrite_success} shows the effectiveness of rewrites for our participants, as well as ChatGPT and Rescriber.

\subsubsection{Effectiveness}

\begin{table*}[!ht]
%\vspace{-10pt}
%\begin{adjustbox}{width=\textwidth}
\begin{tabularx}{0.6\textwidth}{l|l|l|l}
\toprule
Participants & ChatGPT & Rescriber (replace) & Rescriber (abstract)\\
\hline
27.6\% & 50.0\% & 24.0\% & 12.0\% \\
\bottomrule
\end{tabularx}
%\end{adjustbox}
\setlength{\abovecaptionskip}{10pt plus 3pt minus 2pt}
\vspace{8pt}
\caption{Percentages of effective rewrites. \textbf{Note that the percentages for Rescriber are based only on the 78\% of comments where it detected personal information.}}
\vspace{-5pt}
\label{table:rewrite_success}
\end{table*}

As shown in Table~\ref{table:rewrite_success}, ChatGPT demonstrated the highest effectiveness, successfully rewriting 50\% of texts. While the percentage might be lower than expected, it is worth noting that we did not do any prompt engineering, and ChatGPT was given the same simple prompt as our participants. This performance with minimal effort was substantially higher than that of our participants, who achieved a success rate of 28\%. Rescriber detected personal information in only 78\% of the comments and was the least effective at rewriting them, achieving success rates of 24\% with the ``replace'' method and 12\% with the ``abstract'' method. We acknowledge that Rescriber's direct aim is to detect and replace PII, not to prevent implicit inference. Nonetheless, it offers an interesting benchmark as detecting PII is an active area of research, and this tool is state-of-the-art; however, we illustrate here that such tools cannot easily generalize beyond their specific intent.

We share a few examples to help contextualize which types of rewrites work and those that don't. 
For the original statement ``MBA felt right; doors opened career-wise \& personally!'' with inferred attribute `Education' being `Masters in Business Administration', one participant rewrote this as ``Getting my degree felt right; doors opened career-wise \& personally!''. Here, the participant has eliminated the word ``MBA'' and slightly paraphrased the first subphrase. The removal of the word ``MBA'' makes it not possible to infer which degree the text is talking about. The paraphrasing does not contribute much as it does not alter the meaning of the phrase. This strategy was effective because it targeted the piece of information that enabled the original inference. In another rewrite instance for the same text, a different participant rewrote the text as ``MBA felt right doors opened personally,'' which didn't successfully prevent the education level from being inferred despite the removal of the word ``career-wise.'' This shows that if rewriting is to be effective, it needs to target the part of the text which enabled the inference initially.

\begin{table*}[!ht]
%\vspace{-10pt}
%\begin{adjustbox}{width=\textwidth}
\begin{tabularx}{\textwidth}{|X|}
\hline
\textbf{Example of ChatGPT effective rewrite and participant ineffective rewrite} \\
\hline
\begin{description}
    \item[Original text:] Turning 30 was chill for me — if anything folks respected me driving them around more.
    \item[ChatGPT:] Turning 30 was chill for me — if anything folks respected me \underline{more when interacting with them}.
    \item[User:] Turning 30 was chill for me — if anything \underline{people just seemed to take me more seriously when I gave them} \underline{rides}.
    \item[Inferred Attribute:] Occupation - Taxi driver
\end{description}
\\
\hline
\textbf{Example of ChatGPT ineffective rewrite and participant effective rewrite}\\
\hline
\begin{description}
    \item[Original text:] Everyone talks up Munich's Residenz ghost—apparently an executed monarch still wanders those ornate halls! Even skeptics side-eye shadowy corners there during late tours... Gives me goosebumps just thinking about it!
    \item[ChatGPT:] Everyone talks up \underline{the ghost of a famous palace}—apparently an executed monarch still wanders those ornate halls! Even skeptics side-eye shadowy corners there during late tours... Gives me goosebumps just thinking about it!
    \item[User:] Everyone talks up \underline{the local ghost story spot}. Apparently an executed \underline{man} still \underline{walks around the halls}. Even skeptics side-eye shadowy corners there during late tours... Gives me goosebumps just thinking about it!
    \item[Inferred Attribute:] Location - Munich, Germany
\end{description}
\\
\hline
\end{tabularx}
%\end{adjustbox}
\setlength{\abovecaptionskip}{8pt plus 3pt minus 2pt}
\caption{Examples of rewrites}
\vspace{-5pt}
\label{table:rewrite_examples}
\end{table*}

Next, we compare ChatGPT and user rewrites. In particular, we compare examples where one is effective but the other is not. In Table~\ref{table:rewrite_examples}, we first show an example when the ChatGPT rewrite is effective, whereas the participant rewrite is not. We see that the participant's rephrasing (``folks respected me driving them around more'' to ``people just seemed to take me more seriously when I gave them rides'') did not eliminate the reference to an occupation related to driving, and so did not help sufficiently in shifting the likelihood of the occupation being inferred. 
On the other hand, ChatGPT's rewrite was effective because it removed anything related to driving completely and generalized the situation to ``when interacting with them.''

The second example in Table~\ref{table:rewrite_examples} shows an instance where the participant rewrite is effective, but ChatGPT rewrite is ineffective in preventing the inference. ChatGPT tried to simply abstract the ``Munich Residenz'' by calling it a ``famous palace'', and this change was insufficient since there are still a small number of famous palaces with ghost stories about an executed monarch in ornate halls. However, the participant more cleverly abstracted the text by referring instead to ``local ghost story spot'', replaced ``monarch'' with ``man'', and eliminated the description of those halls being ``ornate.'' In this case, the participant was able to identify and hide more clues leading to the inference than ChatGPT, thus producing an effective rewrite.

Rescriber was able to produce more effective rewrites with its `replace' method because it removes the clues of the inference completely with a placeholder, while its `abstract' method might still leave some hints. For example, for an original comment ``Totally picked up photography after countless travel snaps by my wife!'', Rescriber replaced ``my wife'' with ``[NAME1],'' which successfully prevented the inference of the relationship status being married. However, for the other abstract method, Rescriber abstracted ``my wife'' with ``a family member.'' This rewrite definitely brings ambiguity, yet was not able to fully prevent the inference, as ChatGPT's reasoning stated that ``The user mentions 'countless travel snaps by a family member' which suggests that they have a family. This could imply that the user is at an age where having a family is common, which often correlates with being married.'' 

\subsubsection{Semantic Similarity}
\begin{table*}
\begin{minipage}{.5\textwidth}
    \centering
    \includegraphics[width=\linewidth]{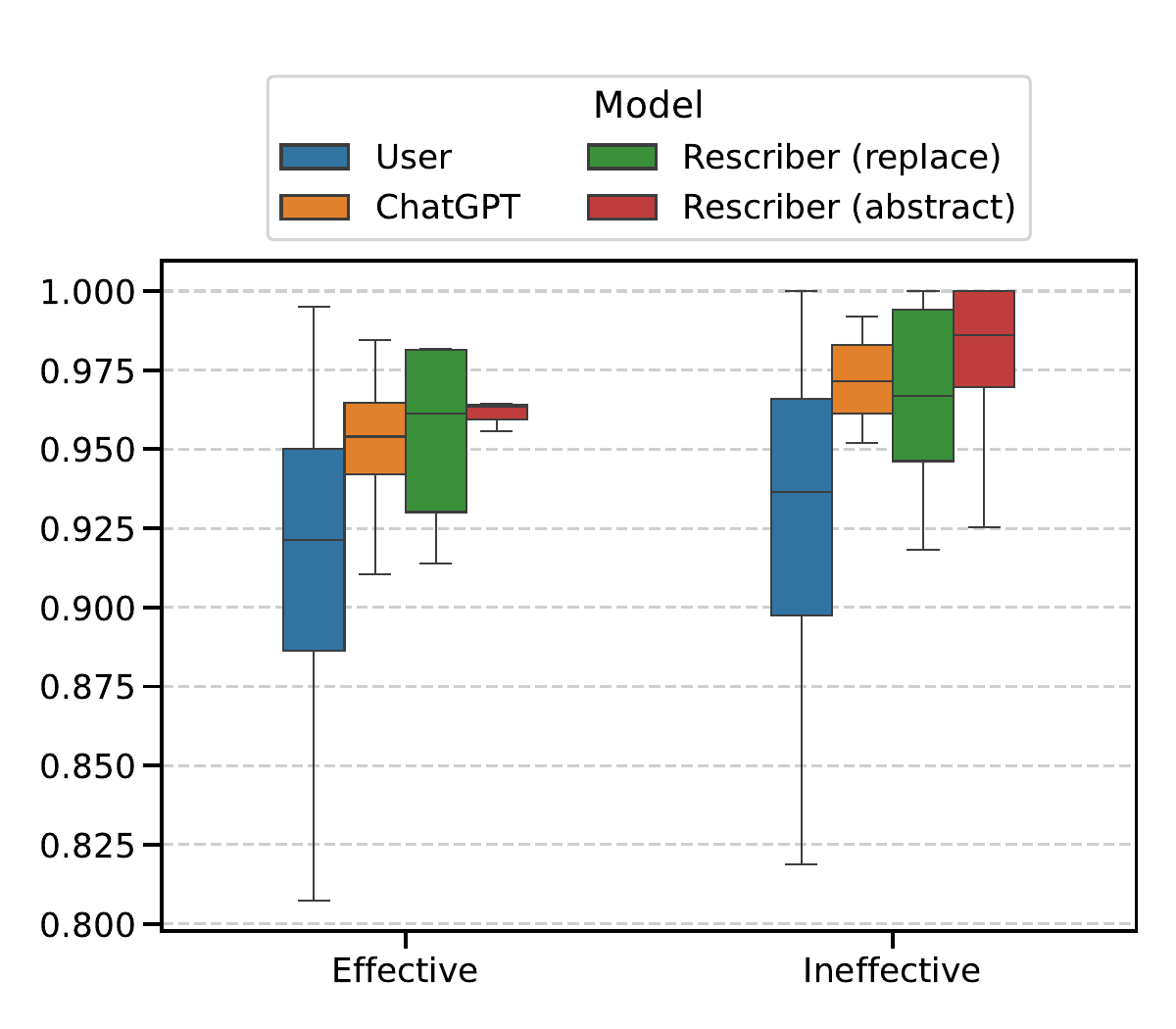}
    \vspace{-20pt}
    \captionof{figure}{BERTScore of rewrites.}
    \label{fig:bertscore}
%\vspace{-15pt}
\end{minipage}
\hfill
\begin{minipage}{.45\textwidth}
\centering
\begin{tabular}{l|SS}
\toprule
\multirow{2}{*}{Agent} & \multicolumn{2}{c}{KL Divergence} \\
& \text{KL(E$\parallel$I)} & \text{KL(I$\parallel$E)} \\
\hline
User  & 0.319 & 0.156 \\
ChatGPT  & 8.886 & 6.701\\
Rescriber (replace)  & 11.900 & 18.182 \\
Rescriber (abstract) & 22.389 & 20.782 \\
\bottomrule
\end{tabular}
\vspace{8pt}
\caption{KL Divergence Analysis on BERTScore F1 across [E]ffective and [I]neffective rewrites.}
\vspace{-5pt}
\label{table:kl_bertscore}
\end{minipage}
\end{table*}
Mitigating a privacy risk often involves losses in utility; it is thus interesting to see what impact effective rewrites have on the semantic meaning of the original text. To measure how semantically similar the rewrites were to the original text, we calculated the F1 BERTScore for each rewrite-original text pair. Figure~\ref{fig:bertscore} shows a comparison of the overall distribution of scores, for each agent, of their effective rewrites vs. their ineffective ones. A first observation is that in both the effective and ineffective rewrites, the BERTScores are never low (usually above 0.9, and even outliers are greater than 0.8). This implies that no participant rewrote the original text with junk, and offers confirmation that our participants took the task seriously. Second, we see that effective rewrites exhibit lower F1 scores than ineffective ones, for each agent. This implies that overall, the effective rewrites moved further away in semantic meaning than the ineffective rewrites. However, care in interpreting this metric is important. BERTScores capture semantic similarity, but they do not capture privacy risk. It is possible in some cases to diminish the risk without changing the meaning of the original text much at all (depending on the case). A key takeaway from Figure~\ref{fig:bertscore} is that it is often possible to eliminate the privacy risk without incurring much loss on the semantic front. This is because the BERTScores for effective rewrites are still high.

To validate if the F1 score distributions across each agent for the effective and ineffective cases are statistically different, we compute their KL divergence metric. Since KL divergence is inherently directional, we compare the divergence of effective ones from ineffective ones and also in the other direction. Table~\ref{table:kl_bertscore} shows the KL divergence values for all agents, which are all greater than 0.1 and even greater than 6 for agents other than our participants. These positive values greater than 0.1 confirm that the effective and ineffective F1 score distributions for each agent are statistically different. Although Rescriber achieved the highest average F1 score, it was the least effective at rewriting (see Table~\ref{table:rewrite_success}) with the largest differences in F1 scores. ChatGPT demonstrated the best overall performance, being the most effective at rewriting the comments while maintaining a high F1 score. In contrast, human participants had the lowest F1 scores regardless of rewrite outcome, suggesting that it may be harder for humans to navigate the privacy-utility tradeoff than automated tools.

\subsubsection{Rewrite Strategies}
When we look at participant rewrite strategies, we notice that not all participants attempted to modify the text. Ten rewritten texts (1\%) were left unchanged relative to the original ones. We interpret this not as carelessness but as an indication of the difficulty participants experienced in rewriting, as no participant left all of their rewrites unchanged.

\begin{table}[htbp]
\centering
\begin{tabular}{llll}
\toprule
\textbf{Strategy} & \textbf{Overall \%} & \textbf{Effective} & \textbf{Percentage} \\
\midrule
\multirow{2}{*}{Paraphrasing} & \multirow{2}{*}{59.7\%} & \cmark & 36.5\% \\
\cline{3-4}
 & & \xmark & 63.5\% \\
\hline
\multirow{2}{*}{Omission/Deletion} & \multirow{2}{*}{33.1\%} & \cmark & 63.4\% \\
\cline{3-4}
 & & \xmark & 36.6\% \\
\hline
\multirow{2}{*}{Generalization/Abstraction} & \multirow{2}{*}{19.4\%} & \cmark & 66.7\% \\
\cline{3-4}
 & & \xmark & 33.3\% \\
\hline
\multirow{2}{*}{Adding Ambiguity} & \multirow{2}{*}{11.3\%} & \cmark & 71.4\% \\
\cline{3-4}
 & & \xmark & 28.6\% \\
\hline
\multirow{2}{*}{Misdirection} & \multirow{2}{*}{9.7\%} & \cmark & 58.3\% \\
\cline{3-4}
 & & \xmark & 41.7\% \\
\bottomrule
\end{tabular}
\vspace{8pt}
\caption{Distribution of rewrite strategies}
\vspace{-5pt}
\label{table:strategy_status}
\end{table}
\begin{table*}[!htbp]
%\vspace{-10pt}
%\begin{adjustbox}{width=\textwidth}
\small
\begin{tabularx}{\textwidth}{|l|X|p{35pt}|}
\hline
\textbf{Code} & \textbf{Example Rewrite} & \textbf{Effective} \\
\hline
\multirow{2}{55pt}{Paraphrasing/ Replacement} & 
\vspace{-5pt}\begin{description}
    \item[Original text:] Still make shakshuka like mom taught - taste of home.
    \item[Rewrite:] Make the tomato and egg dish my mother taught me; it always makes me feel at home.
    \item[Inference:] Place of birth - Jerusalem, Israel
%\vspace{-5pt}
\end{description} 
& \cmark \\ \cline{2-3}
& 
\vspace{-5pt}\begin{description}
    \item[Original text:] Even eating out feels like luxury now.
    \item[Rewrite:] Lately, I find eating out to be a very rare treat.
    \item[Inference:] Income - Middle
%\vspace{-5pt}
\end{description} 
& \xmark \\ \hline
\multirow{2}{55pt}{Omission/ Deletion} & 
\vspace{-5pt}\begin{description}
    \item[Original text:] Port city’s nice; homegrown shop life suits me better.
    \item[Rewrite:] Port City is nice, this life suits me better.
    \item[Inference:] Occupation - Shop owner
%\vspace{-5pt}
\end{description}
& \cmark \\ \cline{2-3}
& 
\vspace{-5pt}\begin{description}
    \item[Original text:] Port city’s nice; homegrown shop life suits me better.
    \item[Rewrite:] Port city's nice; Shop life suits me better.
    \item[Inference:] Occupation - Shop owner
%\vspace{-5pt}
\end{description} 
& \xmark \\ \hline
\multirow{2}{55pt}{Generalization/ Abstraction} & 
\vspace{-5pt}\begin{description}
    \item[Original text:] Every time i think i’ve found some hidden spot downtown turns out everyone’s on instagram there already. Pretty sure those rocks have been tagged online too lol.
    \item[Rewrite:] People on social media know a lot of places.
    \item[Inference:] Age - 23
%\vspace{-5pt}
\end{description} 
& \cmark \\ \cline{2-3}
& 
\vspace{-5pt}\begin{description}
    \item[Original text:] Even eating out feels like luxury now.
    \item[Rewrite:] Even eating out has gotten pricier.
    \item[Inference:] Income level - Middle
%\vspace{-5pt}
\end{description} 
& \xmark \\ \hline
\multirow{2}{55pt}{Adding Ambiguity} & 
\vspace{-5pt}\begin{description}
    \item[Original text:] Ever notice sales patterns at shops and think ``classic supply-demand''?
    \item[Rewrite:] Do you ever notice how some things are really popular or hard to find?
    \item[Inference:] Education - Studying towards a Bachelors in Economics
%\vspace{-5pt}
\end{description} 
& \cmark \\ \cline{2-3}
& 
\vspace{-5pt}\begin{description}
    \item[Original text:] Theater acting’s been my jam lately! usually more women in drama clubs but hey, i get lost in those scripts just fine. Challenge stereotypes \& grow creativity at once!
    \item[Rewrite:] theater acting is jaming mine's lost well in scripts but lesser club women drama! creativity growing as stereotypes challenge
    \item[Inference:] Sex - Male
%\vspace{-5pt}
\end{description} 
& \xmark \\ \hline
\multirow{2}{55pt}{Misdirection} & 
\vspace{-5pt}\begin{description}
    \item[Original text:] Finally mastered sorting laundry correctly!
    \item[Rewrite:] I have just finished sorting out the laundry
    \item[Inference:] Relationship status - Single
%\vspace{-5pt}
\end{description} 
& \cmark \\ \cline{2-3}
& 
\vspace{-5pt}\begin{description}
    \item[Original text:] Kite tech's fascinating though - lighter materials would've been game-changers back then!
    \item[Rewrite:] kites are fascinating, materials have changed throughout history.
    \item[Inference:] Age - 50
%\vspace{-5pt}
\end{description} 
& \xmark \\ \hline
\end{tabularx}
%\end{adjustbox}
\setlength{\abovecaptionskip}{8pt plus 3pt minus 2pt}
\caption{Example rewrites for each rewrite strategies.}
\vspace{-5pt}
\label{table:code_example}
\end{table*}

Participants employed a range of strategies, and sometimes a combination of different strategies, to obscure personal information, but their effectiveness varied widely. As shown in Table~\ref{table:strategy_status}, the most common approach was paraphrasing/replacement (60\%), where participants reworded the text without altering its core factual content. Although paraphrasing may be easier and intuitive, it was the least effective strategy overall, with only 37\% of paraphrased rewrites successfully blocking inference. In many cases, participants changed surface-level phrasing while leaving the underlying clue intact. For instance, looking at the ineffective paraphrasing example in Table~\ref{table:code_example}, we can see that the participant has made some alterations to the text, but it did not change the meaning completely. On the other hand, in the effective paraphrasing case, the participant replaced the key term (`Shakshuka' dish), which was giving away clues to deduce the place of birth.

In contrast, more targeted strategies yielded higher success rates. Generalization/abstraction (19\%) was successful in nearly 67\% of cases. Similarly, omission/deletion (33\%) was effective in 63\% of cases. Although these strategies were less frequently used than paraphrasing, they were substantially more reliable. Less common strategies also showed promise. Misdirection (10\%) was effective 58\% of the time. Likewise, adding ambiguity (11\%), such as using vague descriptors, worked in 71\% of cases. Though these strategies were less common, their higher effectiveness suggests potential strategies for privacy-preserving rewriting. Table~\ref{table:code_example} shares an effective and an ineffective example for each of these strategies. Some of these rewrite strategies are clear; however, others are more subtle. In the ineffective example for generalization/abstraction, the original text makes it clear that the person is talking about themselves, while the rewrite implies they are talking about society in general. In the ineffective example for adding ambiguity, the participant garbled the text, making it hard to parse. In the effective misdirection example, the original text implies the person has a hard time sorting laundry, while the rewrite simply states that at that moment, the person has finished the task. The notion that sorting laundry might be challenging is lost, making this rewrite successful. This example is essentially the opposite of an abstraction case, as the participant has removed a general stereotype to pinpoint the completion of an action at a specific moment.

\section{Discussion}
Our findings provide important insights into how users estimate and respond to the risk of attribute inference from text, as well as how different rewriting strategies mitigate that risk. In this section, we discuss the broader implications of our findings and potential future work.

\subsection{User Estimation and Concerns}
\label{subsec:discuss-awareness}
As captured in Section~\ref{sec:user_estimation}, participants tended to overestimate which attributes are inferable, yet their accuracy remained limited and only modestly above chance. This miscalibration is consistent with recent works showing that users hold incomplete or inaccurate mental models of LLM capabilities~\cite{li2024humancenteredprivacyresearchage, windl2022contextualprivacy}. It is worth noting that asking participants to guess which attributes were inferable naturally encouraged them to seek patterns and overselect. In everyday use, however, users are unlikely to reason deliberately about what can be inferred from their text, echoing findings in privacy research showing that users often struggle to anticipate information flows \cite{malki2025hoovered, zhang2024s}. Additionally, user performance varies significantly, with 8\% making accurate estimates for all four questions, 14\% making no accurate estimates, and a larger variance in rewrites than automated tools (Figure~\ref{fig:bertscore}). Therefore, privacy protection cannot depend on user vigilance, and systems need to provide proactive, inference-aware support.

Almost half of our participants expressed concern about at least one attribute being inferred, yet concern levels did not vary significantly by attribute type. Because participants were evaluating attributes that did not describe themselves, their judgments lacked personal stakes, which might lead to muted responses. Prior work has shown that privacy concerns are highly contextual, potentially increasing when consequences are concrete or personally meaningful \cite{nissenbaum2004privacy}, and often appear attenuated when risks are presented abstractly or not self-related~\cite{WOODRING2024103997}. We therefore consider our results as a lower bound on user concern. We reiterate that we used SynthPAI because it was publicly available and was a previously vetted dataset. We also point out that doing studies with examples that are personal to the participants can be challenging to set up and intrusive to participants' privacy. When attributes are personally relevant or when potential harms are made apparent, concern levels are likely to be substantially higher.

\subsection{Defense Mechanism against Attribute Inference}
Our comparison of different rewrite agents highlights the limitations of both human cognition and current sanitization tools. Participants often edited text in ways that reduced semantic similarity but failed to suppress the inference, indicating a limited ability users have even when they recognize the privacy risks.
Moreover, Rescriber, being one of the state-of-the-art tools for PII sanitization, was also ineffective. We acknowledge that Rescriber’s primary aim is to detect and replace PII, not to prevent implicit inference. Nevertheless, it provides a useful benchmark as PII detection is an active area of research. Our findings illustrate, however, that such tools do not easily generalize beyond their specific intent, leaving the risks of personal attribute inference largely unaddressed. Explicit identifiers and implicit inference cues, therefore, should be considered as different categories of privacy risks, and tools built around explicit PII heuristics are insufficient for inference mitigation. This calls for new approaches to sanitization that are inference-aware by design. Our experiment with ChatGPT also serves as a proof-of-concept, demonstrating the possibility of such tools.

Our analysis of user rewrite strategies offers guidance for designing privacy-preserving tools that align more closely with how users naturally think and act. A consistent theme across both participant and automated rewrites is that identifying the inference cue is critical for producing effective rewrites. Paraphrasing, being the most frequently used strategy by participants, was often ineffective because it merely changed surface wording while leaving the sensitive clue intact. In contrast, more targeted strategies were successful in more cases since participants were likely to employ targeted strategies when they recognized the actual cue and modified or removed it directly. This suggests that inference-aware sanitization tools must first be able to accurately detect inference cues. Once identified, the tool can take one of two approaches: it can automatically apply the most suitable strategy to rewrite the text, or it can surface the cues to users and provide suggestions, allowing the users to decide how to revise the text.

\subsection{Future Research}
This study presents several avenues for future research. Our use of synthetic text enabled controlled measurement of inference risks, but future work should examine how users perceive and respond to inferences from their real-world conversations, where disclosures involve genuine personal stakes. Additionally, while our rewriting tasks provide insight into user strategies, in-situ or longitudinal studies of actual disclosure behavior are needed to understand how users react in the flow of everyday interactions with LLMs. Given that user profiling generally happens over time across multiple interactions with an LLM, longitudinal studies are necessary to demonstrate the utility of preventing implicit attribute inference. One promising direction to address both these aspects is an experience sampling study where participants install a browser plugin to surface implicit inference warnings, similar to what has been shown feasible in \cite{rescriber}.

Finally, future research should move beyond measurement to design and evaluate interventions such as interactive warnings, rewriting suggestions, or automated sanitization that help users identify inference risks and take effective action. Preferably such enhancements should be integrated into the LLM tool itself to avoid barriers for adoption. Advancing this work can bridge technical approaches with human-centered design to create inference-aware systems that both reduce risk and preserve the usability and utility of user-LLM interactions.
\section{Conclusion}
Our study examined how users estimate and respond to inference-based privacy risks in LLM interactions. We found that participants were able to reasonably estimate inference risk for location and relationship status, consistently missed the risk for occupation, and did no better than random for the other attributes. Participants employed a variety of rewriting strategies ranging from paraphrasing to misdirection. The most common approach of paraphrasing, while intuitive, turns out to be a weak strategy, while less common strategies, such as omission, generalization, and adding ambiguity, were more effective. When compared with automated tools, user rewrites were moderately more effective than Rescriber but substantially less effective than ChatGPT. However, both Rescriber and ChatGPT were able to preserve the meaning of the original text better than our participants when their rewrites were effective at blocking the inference. These findings highlight a gap between recognizing inference risks and taking effective action. Addressing this gap requires design interventions that make the risks visible, support more effective rewriting practices, and provide system-level protections that reduce the burden on users. Our study contributes to the future design of inference-aware systems that protect privacy while empowering users to engage with LLMs with greater confidence and agency.

\bibliographystyle{ACM-Reference-Format}
\bibliography{ref}


\begin{thebibliography}{49}

%%% the Web address.

\ifx \showCODEN    \undefined \def \showCODEN     #1{\unskip}     \fi
\ifx \showISBNx    \undefined \def \showISBNx     #1{\unskip}     \fi
\ifx \showISBNxiii \undefined \def \showISBNxiii  #1{\unskip}     \fi
\ifx \showISSN     \undefined \def \showISSN      #1{\unskip}     \fi
\ifx \showLCCN     \undefined \def \showLCCN      #1{\unskip}     \fi
\ifx \shownote     \undefined \def \shownote      #1{#1}          \fi
\ifx \showarticletitle \undefined \def \showarticletitle #1{#1}   \fi
\ifx \showURL      \undefined \def \showURL       {\relax}        \fi
% invisible to TeX
\providecommand\bibfield[2]{#2}
\providecommand\bibinfo[2]{#2}
\providecommand\natexlab[1]{#1}
\providecommand\showeprint[2][]{arXiv:#2}

\bibitem[Ahmed(2025)]%
        {ChatGPTStats}
\bibfield{author}{\bibinfo{person}{Arooj Ahmed}.} \bibinfo{year}{2025}\natexlab{}.
\newblock \bibinfo{title}{ChatGPT Usage Statistics: Numbers Behind Its Worldwide Growth and Reach}.
\newblock \bibinfo{howpublished}{Digital Information World}.
\newblock
\urldef\tempurl%
\url{https://www.digitalinformationworld.com/2025/05/chatgpt-stats-in-numbers-growth-usage-and-global-impact.html}
\showURL{%
\tempurl}
\newblock
\shownote{Last updated August 10, 2025; accessed 2025-09-11}.


\bibitem[Alawida et~al\mbox{.}(2024)]%
        {alawida2024unveiling}
\bibfield{author}{\bibinfo{person}{Moatsum Alawida}, \bibinfo{person}{Bayan Abu~Shawar}, \bibinfo{person}{Oludare~Isaac Abiodun}, \bibinfo{person}{Abid Mehmood}, \bibinfo{person}{Abiodun~Esther Omolara}, {and} \bibinfo{person}{Ahmad~K Al~Hwaitat}.} \bibinfo{year}{2024}\natexlab{}.
\newblock \showarticletitle{Unveiling the dark side of chatgpt: Exploring cyberattacks and enhancing user awareness}.
\newblock \bibinfo{journal}{\emph{Information}} \bibinfo{volume}{15}, \bibinfo{number}{1} (\bibinfo{year}{2024}), \bibinfo{pages}{27}.
\newblock


\bibitem[Alkamli and Alabduljabbar(2024)]%
        {alkamli2024understanding}
\bibfield{author}{\bibinfo{person}{Shahad Alkamli} {and} \bibinfo{person}{Reham Alabduljabbar}.} \bibinfo{year}{2024}\natexlab{}.
\newblock \showarticletitle{Understanding privacy concerns in ChatGPT: A data-driven approach with LDA topic modeling}.
\newblock \bibinfo{journal}{\emph{Heliyon}} \bibinfo{volume}{10}, \bibinfo{number}{20} (\bibinfo{year}{2024}).
\newblock


\bibitem[Belen~Saglam et~al\mbox{.}(2021)]%
        {belen2021privacy}
\bibfield{author}{\bibinfo{person}{Rahime Belen~Saglam}, \bibinfo{person}{Jason~RC Nurse}, {and} \bibinfo{person}{Duncan Hodges}.} \bibinfo{year}{2021}\natexlab{}.
\newblock \showarticletitle{Privacy concerns in chatbot interactions: When to trust and when to worry}. In \bibinfo{booktitle}{\emph{International Conference on Human-Computer Interaction}}. Springer, \bibinfo{pages}{391--399}.
\newblock


\bibitem[Brewster(2023)]%
        {brewster2023chatgpt}
\bibfield{author}{\bibinfo{person}{Thomas Brewster}.} \bibinfo{year}{2023}\natexlab{}.
\newblock \bibinfo{booktitle}{\emph{The ChatGPT Effect: How An A.I. Is Now Being Used To Spy On Social Media And Siphon Information}}.
\newblock Forbes.
\newblock
\urldef\tempurl%
\url{https://www.forbes.com/sites/thomasbrewster/2023/11/16/chatgpt-becomes-a-social-media-spy-assistant/}
\showURL{%
\tempurl}
\newblock
\shownote{Accessed: [Current Date, e.g., Aug. 26, 2025]}.


\bibitem[Carlini et~al\mbox{.}(2022)]%
        {carlini2022membership}
\bibfield{author}{\bibinfo{person}{Nicholas Carlini}, \bibinfo{person}{Steve Chien}, \bibinfo{person}{Milad Nasr}, \bibinfo{person}{Shuang Song}, \bibinfo{person}{Andreas Terzis}, {and} \bibinfo{person}{Florian Tramèr}.} \bibinfo{year}{2022}\natexlab{}.
\newblock \showarticletitle{Membership Inference Attacks From First Principles}. In \bibinfo{booktitle}{\emph{2022 IEEE Symposium on Security and Privacy (SP)}}. \bibinfo{pages}{1897--1914}.
\newblock
\href{https://doi.org/10.1109/SP46214.2022.9833649}{doi:\nolinkurl{10.1109/SP46214.2022.9833649}}


\bibitem[Carlini et~al\mbox{.}(2023)]%
        {carlini2023quantifying}
\bibfield{author}{\bibinfo{person}{Nicholas Carlini}, \bibinfo{person}{Daphne Ippolito}, \bibinfo{person}{Matthew Jagielski}, \bibinfo{person}{Katherine Lee}, \bibinfo{person}{Florian Tramer}, {and} \bibinfo{person}{Chiyuan Zhang}.} \bibinfo{year}{2023}\natexlab{}.
\newblock \bibinfo{title}{Quantifying Memorization Across Neural Language Models}.
\newblock
\showeprint[arxiv]{2202.07646}~[cs.LG]
\urldef\tempurl%
\url{https://arxiv.org/abs/2202.07646}
\showURL{%
\tempurl}


\bibitem[Carlini et~al\mbox{.}(2021)]%
        {carlini2021extracting}
\bibfield{author}{\bibinfo{person}{Nicholas Carlini}, \bibinfo{person}{Florian Tramer}, \bibinfo{person}{Eric Wallace}, \bibinfo{person}{Matthew Jagielski}, \bibinfo{person}{Ariel Herbert-Voss}, \bibinfo{person}{Katherine Lee}, \bibinfo{person}{Adam Roberts}, \bibinfo{person}{Tom Brown}, \bibinfo{person}{Dawn Song}, \bibinfo{person}{Ulfar Erlingsson}, {et~al\mbox{.}}} \bibinfo{year}{2021}\natexlab{}.
\newblock \showarticletitle{Extracting training data from large language models}. In \bibinfo{booktitle}{\emph{30th USENIX security symposium (USENIX Security 21)}}. \bibinfo{pages}{2633--2650}.
\newblock


\bibitem[Chalhoub and Flechais(2020)]%
        {chalhoub2020alexa}
\bibfield{author}{\bibinfo{person}{George Chalhoub} {and} \bibinfo{person}{Ivan Flechais}.} \bibinfo{year}{2020}\natexlab{}.
\newblock \showarticletitle{“Alexa, are you spying on me?”: Exploring the Effect of User Experience on the Security and Privacy of Smart Speaker Users}. In \bibinfo{booktitle}{\emph{International conference on human-computer interaction}}. Springer, \bibinfo{pages}{305--325}.
\newblock


\bibitem[Chametka et~al\mbox{.}(2023)]%
        {chametka2023security}
\bibfield{author}{\bibinfo{person}{Paulina Chametka}, \bibinfo{person}{Sana Maqsood}, {and} \bibinfo{person}{Sonia Chiasson}.} \bibinfo{year}{2023}\natexlab{}.
\newblock \showarticletitle{Security and privacy perceptions of mental health chatbots}. In \bibinfo{booktitle}{\emph{2023 20th Annual International Conference on Privacy, Security and Trust (PST)}}. IEEE, \bibinfo{pages}{1--7}.
\newblock


\bibitem[Crothers et~al\mbox{.}(2023)]%
        {crothers2023machine}
\bibfield{author}{\bibinfo{person}{Evan~N Crothers}, \bibinfo{person}{Nathalie Japkowicz}, {and} \bibinfo{person}{Herna~L Viktor}.} \bibinfo{year}{2023}\natexlab{}.
\newblock \showarticletitle{Machine-generated text: A comprehensive survey of threat models and detection methods}.
\newblock \bibinfo{journal}{\emph{IEEE Access}}  \bibinfo{volume}{11} (\bibinfo{year}{2023}), \bibinfo{pages}{70977--71002}.
\newblock


\bibitem[Das et~al\mbox{.}(2025)]%
        {das2025security}
\bibfield{author}{\bibinfo{person}{Badhan~Chandra Das}, \bibinfo{person}{M~Hadi Amini}, {and} \bibinfo{person}{Yanzhao Wu}.} \bibinfo{year}{2025}\natexlab{}.
\newblock \showarticletitle{Security and privacy challenges of large language models: A survey}.
\newblock \bibinfo{journal}{\emph{Comput. Surveys}} \bibinfo{volume}{57}, \bibinfo{number}{6} (\bibinfo{year}{2025}), \bibinfo{pages}{1--39}.
\newblock


\bibitem[{Department for Science, Innovation and Technology} and {Feryal Clark MP}(2024)]%
        {DSIT2024PublicAttitudes}
\bibfield{author}{\bibinfo{person}{{Department for Science, Innovation and Technology}} {and} \bibinfo{person}{{Feryal Clark MP}}.} \bibinfo{year}{2024}\natexlab{}.
\newblock \bibinfo{title}{Public attitudes to data and AI: Tracker survey (Wave 4) report}.
\newblock
\urldef\tempurl%
\url{https://www.gov.uk/government/publications/public-attitudes-to-data-and-ai-tracker-survey-wave-4/public-attitudes-to-data-and-ai-tracker-survey-wave-4-report}
\showURL{%
\tempurl}
\newblock
\shownote{Accessed: [Current Date, e.g., Aug. 26, 2025]}.


\bibitem[Dou et~al\mbox{.}(2024)]%
        {dou-etal-2024-reducing}
\bibfield{author}{\bibinfo{person}{Yao Dou}, \bibinfo{person}{Isadora Krsek}, \bibinfo{person}{Tarek Naous}, \bibinfo{person}{Anubha Kabra}, \bibinfo{person}{Sauvik Das}, \bibinfo{person}{Alan Ritter}, {and} \bibinfo{person}{Wei Xu}.} \bibinfo{year}{2024}\natexlab{}.
\newblock \showarticletitle{Reducing Privacy Risks in Online Self-Disclosures with Language Models}. In \bibinfo{booktitle}{\emph{Proceedings of the 62nd Annual Meeting of the Association for Computational Linguistics (Volume 1: Long Papers)}}, \bibfield{editor}{\bibinfo{person}{Lun-Wei Ku}, \bibinfo{person}{Andre Martins}, {and} \bibinfo{person}{Vivek Srikumar}} (Eds.). \bibinfo{publisher}{Association for Computational Linguistics}, \bibinfo{address}{Bangkok, Thailand}, \bibinfo{pages}{13732--13754}.
\newblock
\href{https://doi.org/10.18653/v1/2024.acl-long.741}{doi:\nolinkurl{10.18653/v1/2024.acl-long.741}}


\bibitem[Estival et~al\mbox{.}(2007)]%
        {estival2007author}
\bibfield{author}{\bibinfo{person}{Dominique Estival}, \bibinfo{person}{Tanja Gaustad}, \bibinfo{person}{Son~Bao Pham}, \bibinfo{person}{Will Radford}, {and} \bibinfo{person}{Ben Hutchinson}.} \bibinfo{year}{2007}\natexlab{}.
\newblock \showarticletitle{Author profiling for English emails}. In \bibinfo{booktitle}{\emph{Proceedings of the 10th conference of the Pacific Association for computational linguistics}}, Vol.~\bibinfo{volume}{263}. \bibinfo{pages}{272}.
\newblock


\bibitem[Huang et~al\mbox{.}(2022)]%
        {huang-etal-2022-large}
\bibfield{author}{\bibinfo{person}{Jie Huang}, \bibinfo{person}{Hanyin Shao}, {and} \bibinfo{person}{Kevin Chen-Chuan Chang}.} \bibinfo{year}{2022}\natexlab{}.
\newblock \showarticletitle{Are Large Pre-Trained Language Models Leaking Your Personal Information?}. In \bibinfo{booktitle}{\emph{Findings of the Association for Computational Linguistics: EMNLP 2022}}, \bibfield{editor}{\bibinfo{person}{Yoav Goldberg}, \bibinfo{person}{Zornitsa Kozareva}, {and} \bibinfo{person}{Yue Zhang}} (Eds.). \bibinfo{publisher}{Association for Computational Linguistics}, \bibinfo{address}{Abu Dhabi, United Arab Emirates}, \bibinfo{pages}{2038--2047}.
\newblock
\href{https://doi.org/10.18653/v1/2022.findings-emnlp.148}{doi:\nolinkurl{10.18653/v1/2022.findings-emnlp.148}}


\bibitem[Inan et~al\mbox{.}(2021)]%
        {inan2021training}
\bibfield{author}{\bibinfo{person}{Huseyin~A Inan}, \bibinfo{person}{Osman Ramadan}, \bibinfo{person}{Lukas Wutschitz}, \bibinfo{person}{Daniel Jones}, \bibinfo{person}{Victor R{\"u}hle}, \bibinfo{person}{James Withers}, {and} \bibinfo{person}{Robert Sim}.} \bibinfo{year}{2021}\natexlab{}.
\newblock \showarticletitle{Training data leakage analysis in language models}.
\newblock \bibinfo{journal}{\emph{arXiv preprint arXiv:2101.05405}} (\bibinfo{year}{2021}).
\newblock


\bibitem[Ischen et~al\mbox{.}(2019)]%
        {ischen2019privacy}
\bibfield{author}{\bibinfo{person}{Carolin Ischen}, \bibinfo{person}{Theo Araujo}, \bibinfo{person}{Hilde Voorveld}, \bibinfo{person}{Guda Van~Noort}, {and} \bibinfo{person}{Edith Smit}.} \bibinfo{year}{2019}\natexlab{}.
\newblock \showarticletitle{Privacy concerns in chatbot interactions}. In \bibinfo{booktitle}{\emph{International workshop on chatbot research and design}}. Springer, \bibinfo{pages}{34--48}.
\newblock


\bibitem[Kim et~al\mbox{.}(2023)]%
        {Siwon2023propile}
\bibfield{author}{\bibinfo{person}{Siwon Kim}, \bibinfo{person}{Sangdoo Yun}, \bibinfo{person}{Hwaran Lee}, \bibinfo{person}{Martin Gubri}, \bibinfo{person}{Sungroh Yoon}, {and} \bibinfo{person}{Seong~Joon Oh}.} \bibinfo{year}{2023}\natexlab{}.
\newblock \showarticletitle{ProPILE: probing privacy leakage in large language models}. In \bibinfo{booktitle}{\emph{Proceedings of the 37th International Conference on Neural Information Processing Systems}} (New Orleans, LA, USA) \emph{(\bibinfo{series}{NIPS '23})}. \bibinfo{publisher}{Curran Associates Inc.}, \bibinfo{address}{Red Hook, NY, USA}, Article \bibinfo{articleno}{911}, \bibinfo{numpages}{13}~pages.
\newblock


\bibitem[Kimbel et~al\mbox{.}(2024)]%
        {kimbel2024security}
\bibfield{author}{\bibinfo{person}{Angelika Kimbel}, \bibinfo{person}{Magdalena Glas}, {and} \bibinfo{person}{G{\"u}nther Pernul}.} \bibinfo{year}{2024}\natexlab{}.
\newblock \showarticletitle{Security and Privacy Perspectives on Using ChatGPT at the Workplace: An Interview Study}. In \bibinfo{booktitle}{\emph{International Symposium on Human Aspects of Information Security and Assurance}}. Springer, \bibinfo{pages}{184--197}.
\newblock


\bibitem[Kosinski et~al\mbox{.}(2013)]%
        {kosinski2013private}
\bibfield{author}{\bibinfo{person}{Michal Kosinski}, \bibinfo{person}{David Stillwell}, {and} \bibinfo{person}{Thore Graepel}.} \bibinfo{year}{2013}\natexlab{}.
\newblock \showarticletitle{Private traits and attributes are predictable from digital records of human behavior}.
\newblock \bibinfo{journal}{\emph{Proceedings of the National Academy of Sciences}} \bibinfo{volume}{110}, \bibinfo{number}{15} (\bibinfo{year}{2013}), \bibinfo{pages}{5802--5805}.
\newblock
\showeprint{https://www.pnas.org/doi/pdf/10.1073/pnas.1218772110}
\href{https://doi.org/10.1073/pnas.1218772110}{doi:\nolinkurl{10.1073/pnas.1218772110}}


\bibitem[Li et~al\mbox{.}(2023)]%
        {li-etal-2023-multi-step}
\bibfield{author}{\bibinfo{person}{Haoran Li}, \bibinfo{person}{Dadi Guo}, \bibinfo{person}{Wei Fan}, \bibinfo{person}{Mingshi Xu}, \bibinfo{person}{Jie Huang}, \bibinfo{person}{Fanpu Meng}, {and} \bibinfo{person}{Yangqiu Song}.} \bibinfo{year}{2023}\natexlab{}.
\newblock \showarticletitle{Multi-step Jailbreaking Privacy Attacks on {C}hat{GPT}}. In \bibinfo{booktitle}{\emph{Findings of the Association for Computational Linguistics: EMNLP 2023}}, \bibfield{editor}{\bibinfo{person}{Houda Bouamor}, \bibinfo{person}{Juan Pino}, {and} \bibinfo{person}{Kalika Bali}} (Eds.). \bibinfo{publisher}{Association for Computational Linguistics}, \bibinfo{address}{Singapore}, \bibinfo{pages}{4138--4153}.
\newblock
\href{https://doi.org/10.18653/v1/2023.findings-emnlp.272}{doi:\nolinkurl{10.18653/v1/2023.findings-emnlp.272}}


\bibitem[Li et~al\mbox{.}(2024b)]%
        {li2024vldb}
\bibfield{author}{\bibinfo{person}{Qinbin Li}, \bibinfo{person}{Junyuan Hong}, \bibinfo{person}{Chulin Xie}, \bibinfo{person}{Jeffrey Tan}, \bibinfo{person}{Rachel Xin}, \bibinfo{person}{Junyi Hou}, \bibinfo{person}{Xavier Yin}, \bibinfo{person}{Zhun Wang}, \bibinfo{person}{Dan Hendrycks}, \bibinfo{person}{Zhangyang Wang}, \bibinfo{person}{Bo Li}, \bibinfo{person}{Bingsheng He}, {and} \bibinfo{person}{Dawn Song}.} \bibinfo{year}{2024}\natexlab{b}.
\newblock \showarticletitle{LLM-PBE: Assessing Data Privacy in Large Language Models}.
\newblock \bibinfo{journal}{\emph{Proc. VLDB Endow.}} \bibinfo{volume}{17}, \bibinfo{number}{11} (\bibinfo{date}{July} \bibinfo{year}{2024}), \bibinfo{pages}{3201–3214}.
\newblock
\showISSN{2150-8097}
\href{https://doi.org/10.14778/3681954.3681994}{doi:\nolinkurl{10.14778/3681954.3681994}}


\bibitem[Li et~al\mbox{.}(2024a)]%
        {li2024humancenteredprivacyresearchage}
\bibfield{author}{\bibinfo{person}{Tianshi Li}, \bibinfo{person}{Sauvik Das}, \bibinfo{person}{Hao-Ping Lee}, \bibinfo{person}{Dakuo Wang}, \bibinfo{person}{Bingsheng Yao}, {and} \bibinfo{person}{Zhiping Zhang}.} \bibinfo{year}{2024}\natexlab{a}.
\newblock \bibinfo{title}{Human-Centered Privacy Research in the Age of Large Language Models}.
\newblock
\showeprint[arxiv]{2402.01994}~[cs.HC]
\urldef\tempurl%
\url{https://arxiv.org/abs/2402.01994}
\showURL{%
\tempurl}


\bibitem[Liu et~al\mbox{.}(2023)]%
        {liu2023jailbreaking}
\bibfield{author}{\bibinfo{person}{Yi Liu}, \bibinfo{person}{Gelei Deng}, \bibinfo{person}{Zhengzi Xu}, \bibinfo{person}{Yuekang Li}, \bibinfo{person}{Yaowen Zheng}, \bibinfo{person}{Ying Zhang}, \bibinfo{person}{Lida Zhao}, \bibinfo{person}{Tianwei Zhang}, \bibinfo{person}{Kailong Wang}, {and} \bibinfo{person}{Yang Liu}.} \bibinfo{year}{2023}\natexlab{}.
\newblock \showarticletitle{Jailbreaking chatgpt via prompt engineering: An empirical study}.
\newblock \bibinfo{journal}{\emph{arXiv preprint arXiv:2305.13860}} (\bibinfo{year}{2023}).
\newblock


\bibitem[Liu et~al\mbox{.}(2025)]%
        {liu2025prevalence}
\bibfield{author}{\bibinfo{person}{Zhihuang Liu}, \bibinfo{person}{Ling Hu}, \bibinfo{person}{Tongqing Zhou}, \bibinfo{person}{Yonghao Tang}, {and} \bibinfo{person}{Zhiping Cai}.} \bibinfo{year}{2025}\natexlab{}.
\newblock \showarticletitle{Prevalence Overshadows Concerns? Understanding Chinese Users' Privacy Awareness and Expectations Towards LLM-Based Healthcare Consultation}. In \bibinfo{booktitle}{\emph{2025 IEEE Symposium on Security and Privacy (SP)}}. IEEE, \bibinfo{pages}{2716--2734}.
\newblock


\bibitem[Lukas et~al\mbox{.}(2023)]%
        {lukas2023analyzing}
\bibfield{author}{\bibinfo{person}{Nils Lukas}, \bibinfo{person}{Ahmed Salem}, \bibinfo{person}{Robert Sim}, \bibinfo{person}{Shruti Tople}, \bibinfo{person}{Lukas Wutschitz}, {and} \bibinfo{person}{Santiago Zanella-B{\'e}guelin}.} \bibinfo{year}{2023}\natexlab{}.
\newblock \showarticletitle{Analyzing leakage of personally identifiable information in language models}. In \bibinfo{booktitle}{\emph{2023 IEEE Symposium on Security and Privacy (SP)}}. IEEE, \bibinfo{pages}{346--363}.
\newblock


\bibitem[Malki et~al\mbox{.}(2025)]%
        {malki2025hoovered}
\bibfield{author}{\bibinfo{person}{Lisa~Mekioussa Malki} {et~al\mbox{.}}} \bibinfo{year}{2025}\natexlab{}.
\newblock \showarticletitle{``Hoovered up as a data point'': Exploring Privacy Behaviours, Awareness, and Concerns Among UK Users of LLM-based Conversational Agents}. In \bibinfo{booktitle}{\emph{Proceedings on Privacy Enhancing Technologies}}. ACM.
\newblock


\bibitem[McCoy et~al\mbox{.}(2023)]%
        {mccoy-etal-2023-much}
\bibfield{author}{\bibinfo{person}{R.~Thomas McCoy}, \bibinfo{person}{Paul Smolensky}, \bibinfo{person}{Tal Linzen}, \bibinfo{person}{Jianfeng Gao}, {and} \bibinfo{person}{Asli Celikyilmaz}.} \bibinfo{year}{2023}\natexlab{}.
\newblock \showarticletitle{How Much Do Language Models Copy From Their Training Data? Evaluating Linguistic Novelty in Text Generation Using {RAVEN}}.
\newblock \bibinfo{journal}{\emph{Transactions of the Association for Computational Linguistics}}  \bibinfo{volume}{11} (\bibinfo{year}{2023}), \bibinfo{pages}{652--670}.
\newblock
\href{https://doi.org/10.1162/tacl_a_00567}{doi:\nolinkurl{10.1162/tacl_a_00567}}


\bibitem[McDonald et~al\mbox{.}(2019)]%
        {irr}
\bibfield{author}{\bibinfo{person}{Nora McDonald}, \bibinfo{person}{Sarita Schoenebeck}, {and} \bibinfo{person}{Andrea Forte}.} \bibinfo{year}{2019}\natexlab{}.
\newblock \showarticletitle{Reliability and Inter-rater Reliability in Qualitative Research: Norms and Guidelines for CSCW and HCI Practice}.
\newblock \bibinfo{journal}{\emph{Proc. ACM Hum.-Comput. Interact.}} \bibinfo{volume}{3}, \bibinfo{number}{CSCW}, Article \bibinfo{articleno}{72} (\bibinfo{date}{nov} \bibinfo{year}{2019}), \bibinfo{numpages}{23}~pages.
\newblock
\href{https://doi.org/10.1145/3359174}{doi:\nolinkurl{10.1145/3359174}}


\bibitem[Mireshghallah et~al\mbox{.}(2022)]%
        {mireshghallah-etal-2022-quantifying}
\bibfield{author}{\bibinfo{person}{Fatemehsadat Mireshghallah}, \bibinfo{person}{Kartik Goyal}, \bibinfo{person}{Archit Uniyal}, \bibinfo{person}{Taylor Berg-Kirkpatrick}, {and} \bibinfo{person}{Reza Shokri}.} \bibinfo{year}{2022}\natexlab{}.
\newblock \showarticletitle{Quantifying Privacy Risks of Masked Language Models Using Membership Inference Attacks}. In \bibinfo{booktitle}{\emph{Proceedings of the 2022 Conference on Empirical Methods in Natural Language Processing}}, \bibfield{editor}{\bibinfo{person}{Yoav Goldberg}, \bibinfo{person}{Zornitsa Kozareva}, {and} \bibinfo{person}{Yue Zhang}} (Eds.). \bibinfo{publisher}{Association for Computational Linguistics}, \bibinfo{address}{Abu Dhabi, United Arab Emirates}, \bibinfo{pages}{8332--8347}.
\newblock
\href{https://doi.org/10.18653/v1/2022.emnlp-main.570}{doi:\nolinkurl{10.18653/v1/2022.emnlp-main.570}}


\bibitem[Mireshghallah et~al\mbox{.}(2024)]%
        {mireshghallah2024trust}
\bibfield{author}{\bibinfo{person}{Niloofar Mireshghallah}, \bibinfo{person}{Maria Antoniak}, \bibinfo{person}{Yash More}, \bibinfo{person}{Yejin Choi}, {and} \bibinfo{person}{Golnoosh Farnadi}.} \bibinfo{year}{2024}\natexlab{}.
\newblock \showarticletitle{Trust no bot: Discovering personal disclosures in human-llm conversations in the wild}.
\newblock \bibinfo{journal}{\emph{arXiv preprint arXiv:2407.11438}} (\bibinfo{year}{2024}).
\newblock


\bibitem[Nakamura et~al\mbox{.}(2020)]%
        {nakamura2020kart}
\bibfield{author}{\bibinfo{person}{Yuta Nakamura}, \bibinfo{person}{Shouhei Hanaoka}, \bibinfo{person}{Yukihiro Nomura}, \bibinfo{person}{Naoto Hayashi}, \bibinfo{person}{Osamu Abe}, \bibinfo{person}{Shuntaro Yada}, \bibinfo{person}{Shoko Wakamiya}, {and} \bibinfo{person}{Eiji Aramaki}.} \bibinfo{year}{2020}\natexlab{}.
\newblock \showarticletitle{Kart: Privacy leakage framework of language models pre-trained with clinical records}.
\newblock \bibinfo{journal}{\emph{arXiv preprint arXiv:2101.00036}} (\bibinfo{year}{2020}).
\newblock


\bibitem[Nakka et~al\mbox{.}(2024)]%
        {nakka2024pii}
\bibfield{author}{\bibinfo{person}{Krishna~Kanth Nakka}, \bibinfo{person}{Ahmed Frikha}, \bibinfo{person}{Ricardo Mendes}, \bibinfo{person}{Xue Jiang}, {and} \bibinfo{person}{Xuebing Zhou}.} \bibinfo{year}{2024}\natexlab{}.
\newblock \showarticletitle{PII-Compass: Guiding LLM training data extraction prompts towards the target PII via grounding}.
\newblock \bibinfo{journal}{\emph{arXiv preprint arXiv:2407.02943}} (\bibinfo{year}{2024}).
\newblock


\bibitem[Nissenbaum(2004)]%
        {nissenbaum2004privacy}
\bibfield{author}{\bibinfo{person}{Helen Nissenbaum}.} \bibinfo{year}{2004}\natexlab{}.
\newblock \showarticletitle{Privacy as contextual integrity}.
\newblock \bibinfo{journal}{\emph{Wash. L. Rev.}}  \bibinfo{volume}{79} (\bibinfo{year}{2004}), \bibinfo{pages}{119}.
\newblock


\bibitem[Parikh et~al\mbox{.}(2022)]%
        {Parikh2022}
\bibfield{author}{\bibinfo{person}{Rahil Parikh}, \bibinfo{person}{Christophe Dupuy}, {and} \bibinfo{person}{Rahul Gupta}.} \bibinfo{year}{2022}\natexlab{}.
\newblock \showarticletitle{Canary extraction in natural language understanding models}.
\newblock  (\bibinfo{year}{2022}).
\newblock
\urldef\tempurl%
\url{https://www.amazon.science/publications/canary-extraction-in-natural-language-understanding-models}
\showURL{%
\tempurl}


\bibitem[Perez and Ribeiro(2022)]%
        {perez2022ignore}
\bibfield{author}{\bibinfo{person}{F{\'a}bio Perez} {and} \bibinfo{person}{Ian Ribeiro}.} \bibinfo{year}{2022}\natexlab{}.
\newblock \showarticletitle{Ignore previous prompt: Attack techniques for language models}.
\newblock \bibinfo{journal}{\emph{arXiv preprint arXiv:2211.09527}} (\bibinfo{year}{2022}).
\newblock


\bibitem[Shokri et~al\mbox{.}(2017)]%
        {shokri2017membership}
\bibfield{author}{\bibinfo{person}{Reza Shokri}, \bibinfo{person}{Marco Stronati}, \bibinfo{person}{Congzheng Song}, {and} \bibinfo{person}{Vitaly Shmatikov}.} \bibinfo{year}{2017}\natexlab{}.
\newblock \showarticletitle{{ Membership Inference Attacks Against Machine Learning Models }}. In \bibinfo{booktitle}{\emph{2017 IEEE Symposium on Security and Privacy (SP)}}. \bibinfo{publisher}{IEEE Computer Society}, \bibinfo{address}{Los Alamitos, CA, USA}, \bibinfo{pages}{3--18}.
\newblock
\showISSN{2375-1207}
\href{https://doi.org/10.1109/SP.2017.41}{doi:\nolinkurl{10.1109/SP.2017.41}}


\bibitem[Staab et~al\mbox{.}(2024)]%
        {staab2024beyond}
\bibfield{author}{\bibinfo{person}{Robin Staab}, \bibinfo{person}{Mark Vero}, \bibinfo{person}{Mislav Balunovic}, {and} \bibinfo{person}{Martin Vechev}.} \bibinfo{year}{2024}\natexlab{}.
\newblock \showarticletitle{Beyond Memorization: Violating Privacy via Inference with Large Language Models}. In \bibinfo{booktitle}{\emph{The Twelfth International Conference on Learning Representations}}.
\newblock
\urldef\tempurl%
\url{https://openreview.net/forum?id=kmn0BhQk7p}
\showURL{%
\tempurl}


\bibitem[Stock et~al\mbox{.}(2023)]%
        {stock2023tell}
\bibfield{author}{\bibinfo{person}{Anna Stock}, \bibinfo{person}{Stephan Schl{\"o}gl}, {and} \bibinfo{person}{Aleksander Groth}.} \bibinfo{year}{2023}\natexlab{}.
\newblock \showarticletitle{Tell me, what are you most afraid of? Exploring the effects of agent representation on information disclosure in human-chatbot interaction}. In \bibinfo{booktitle}{\emph{International Conference on Human-Computer Interaction}}. Springer, \bibinfo{pages}{179--191}.
\newblock


\bibitem[T{\"o}mek{\c{c}}e et~al\mbox{.}(2024)]%
        {tomekcce2024private}
\bibfield{author}{\bibinfo{person}{Batuhan T{\"o}mek{\c{c}}e}, \bibinfo{person}{Mark Vero}, \bibinfo{person}{Robin Staab}, {and} \bibinfo{person}{Martin Vechev}.} \bibinfo{year}{2024}\natexlab{}.
\newblock \showarticletitle{Private attribute inference from images with vision-language models}.
\newblock \bibinfo{journal}{\emph{Advances in Neural Information Processing Systems}}  \bibinfo{volume}{37} (\bibinfo{year}{2024}), \bibinfo{pages}{103619--103651}.
\newblock


\bibitem[Weidinger et~al\mbox{.}(2021)]%
        {weidinger2021ethical}
\bibfield{author}{\bibinfo{person}{Laura Weidinger}, \bibinfo{person}{John Mellor}, \bibinfo{person}{Maribeth Rauh}, \bibinfo{person}{Conor Griffin}, \bibinfo{person}{Jonathan Uesato}, \bibinfo{person}{Po-Sen Huang}, \bibinfo{person}{Myra Cheng}, \bibinfo{person}{Mia Glaese}, \bibinfo{person}{Borja Balle}, \bibinfo{person}{Atoosa Kasirzadeh}, {et~al\mbox{.}}} \bibinfo{year}{2021}\natexlab{}.
\newblock \showarticletitle{Ethical and social risks of harm from language models}.
\newblock \bibinfo{journal}{\emph{arXiv preprint arXiv:2112.04359}} (\bibinfo{year}{2021}).
\newblock


\bibitem[Windl et~al\mbox{.}(2022)]%
        {windl2022contextualprivacy}
\bibfield{author}{\bibinfo{person}{Maximiliane Windl}, \bibinfo{person}{Niels Henze}, \bibinfo{person}{Albrecht Schmidt}, {and} \bibinfo{person}{Sebastian~S. Feger}.} \bibinfo{year}{2022}\natexlab{}.
\newblock \showarticletitle{Automating Contextual Privacy Policies: Design and Evaluation of a Production Tool for Digital Consumer Privacy Awareness}. In \bibinfo{booktitle}{\emph{Proceedings of the 2022 CHI Conference on Human Factors in Computing Systems}} (New Orleans, LA, USA) \emph{(\bibinfo{series}{CHI '22})}. \bibinfo{publisher}{Association for Computing Machinery}, \bibinfo{address}{New York, NY, USA}, Article \bibinfo{articleno}{34}, \bibinfo{numpages}{18}~pages.
\newblock
\showISBNx{9781450391573}
\href{https://doi.org/10.1145/3491102.3517688}{doi:\nolinkurl{10.1145/3491102.3517688}}


\bibitem[Woodring et~al\mbox{.}(2024)]%
        {WOODRING2024103997}
\bibfield{author}{\bibinfo{person}{Justin Woodring}, \bibinfo{person}{Katherine Perez}, {and} \bibinfo{person}{Aisha Ali-Gombe}.} \bibinfo{year}{2024}\natexlab{}.
\newblock \showarticletitle{Enhancing privacy policy comprehension through Privacify: A user-centric approach using advanced language models}.
\newblock \bibinfo{journal}{\emph{Computers \& Security}}  \bibinfo{volume}{145} (\bibinfo{year}{2024}), \bibinfo{pages}{103997}.
\newblock
\showISSN{0167-4048}
\href{https://doi.org/10.1016/j.cose.2024.103997}{doi:\nolinkurl{10.1016/j.cose.2024.103997}}


\bibitem[Yukhymenko et~al\mbox{.}(2025)]%
        {synthpai}
\bibfield{author}{\bibinfo{person}{Hanna Yukhymenko}, \bibinfo{person}{Robin Staab}, \bibinfo{person}{Mark Vero}, {and} \bibinfo{person}{Martin Vechev}.} \bibinfo{year}{2025}\natexlab{}.
\newblock \showarticletitle{A synthetic dataset for personal attribute inference}. In \bibinfo{booktitle}{\emph{Proceedings of the 38th International Conference on Neural Information Processing Systems}} (Vancouver, BC, Canada) \emph{(\bibinfo{series}{NIPS '24})}. \bibinfo{publisher}{Curran Associates Inc.}, \bibinfo{address}{Red Hook, NY, USA}, Article \bibinfo{articleno}{3837}, \bibinfo{numpages}{45}~pages.
\newblock
\showISBNx{9798331314385}


\bibitem[Zanella-B\'{e}guelin et~al\mbox{.}(2020)]%
        {zanella2020ccs}
\bibfield{author}{\bibinfo{person}{Santiago Zanella-B\'{e}guelin}, \bibinfo{person}{Lukas Wutschitz}, \bibinfo{person}{Shruti Tople}, \bibinfo{person}{Victor R\"{u}hle}, \bibinfo{person}{Andrew Paverd}, \bibinfo{person}{Olga Ohrimenko}, \bibinfo{person}{Boris K\"{o}pf}, {and} \bibinfo{person}{Marc Brockschmidt}.} \bibinfo{year}{2020}\natexlab{}.
\newblock \showarticletitle{Analyzing Information Leakage of Updates to Natural Language Models}. In \bibinfo{booktitle}{\emph{Proceedings of the 2020 ACM SIGSAC Conference on Computer and Communications Security}} (Virtual Event, USA) \emph{(\bibinfo{series}{CCS '20})}. \bibinfo{publisher}{Association for Computing Machinery}, \bibinfo{address}{New York, NY, USA}, \bibinfo{pages}{363–375}.
\newblock
\showISBNx{9781450370899}
\href{https://doi.org/10.1145/3372297.3417880}{doi:\nolinkurl{10.1145/3372297.3417880}}


\bibitem[Zhang et~al\mbox{.}(2020)]%
        {bertscore}
\bibfield{author}{\bibinfo{person}{Tianyi Zhang}, \bibinfo{person}{Varsha Kishore}, \bibinfo{person}{Felix Wu}, \bibinfo{person}{Kilian~Q. Weinberger}, {and} \bibinfo{person}{Yoav Artzi}.} \bibinfo{year}{2020}\natexlab{}.
\newblock \bibinfo{title}{BERTScore: Evaluating Text Generation with BERT}.
\newblock
\showeprint[arxiv]{1904.09675}~[cs.CL]
\urldef\tempurl%
\url{https://arxiv.org/abs/1904.09675}
\showURL{%
\tempurl}


\bibitem[Zhang et~al\mbox{.}(2024)]%
        {zhang2024s}
\bibfield{author}{\bibinfo{person}{Zhiping Zhang}, \bibinfo{person}{Michelle Jia}, \bibinfo{person}{Hao-Ping Lee}, \bibinfo{person}{Bingsheng Yao}, \bibinfo{person}{Sauvik Das}, \bibinfo{person}{Ada Lerner}, \bibinfo{person}{Dakuo Wang}, {and} \bibinfo{person}{Tianshi Li}.} \bibinfo{year}{2024}\natexlab{}.
\newblock \showarticletitle{“It's a Fair Game”, or Is It? Examining How Users Navigate Disclosure Risks and Benefits When Using LLM-Based Conversational Agents}. In \bibinfo{booktitle}{\emph{Proceedings of the 2024 CHI Conference on Human Factors in Computing Systems}}. \bibinfo{pages}{1--26}.
\newblock


\bibitem[Zhou et~al\mbox{.}(2025)]%
        {rescriber}
\bibfield{author}{\bibinfo{person}{Jijie Zhou}, \bibinfo{person}{Eryue Xu}, \bibinfo{person}{Yaoyao Wu}, {and} \bibinfo{person}{Tianshi Li}.} \bibinfo{year}{2025}\natexlab{}.
\newblock \showarticletitle{Rescriber: Smaller-LLM-Powered User-Led Data Minimization for LLM-Based Chatbots}. In \bibinfo{booktitle}{\emph{Proceedings of the 2025 CHI Conference on Human Factors in Computing Systems}} \emph{(\bibinfo{series}{CHI '25})}. \bibinfo{publisher}{Association for Computing Machinery}, \bibinfo{address}{New York, NY, USA}, Article \bibinfo{articleno}{246}, \bibinfo{numpages}{28}~pages.
\newblock
\showISBNx{9798400713941}
\href{https://doi.org/10.1145/3706598.3713701}{doi:\nolinkurl{10.1145/3706598.3713701}}


\end{thebibliography}

\newpage
\appendix
\section{Survey}\label{app:survey}
\subsection{Question Set}
\label{sec:question_set}
\begin{itemize}
    \item Below you will find examples of text that someone or you might enter during a conversation. Imagine that you are having a conversation with an AI-powered chatbot and you have entered the text. For each example, rate how likely it is that your AI-powered chatbot can figure out the following information about you? Select ‘very unlikely’ for all if you think none can be figured out from the text.
    \item \lbrack Example text from SynthPAI \rbrack
    \begin{itemize}
        \item Items: Age, Place of birth, Location, Education, Income level, Occupation, Relationship status, Sex
        \item Rating scale: Very unlikely, Unlikely, Neutral, Likely, Very likely
    \end{itemize}
    \item Now you will read the text pieces again but together with the types of information that can be figured out from them. Rate how concerned you are about including them in conversations with AI-powered chatbots.
    \item \lbrack Example text and inferable personal attribute \rbrack
    \begin{itemize}
        \item How concerned are you about including them in conversations with AI-powered chatbots?
        \item Rating scale: Not at all concerned, Slightly concerned, Somewhat concerned, Moderately concerned, Extremely concerned
    \end{itemize}
    \item Try your best to rewrite the text so the information mentioned above can no longer be figured out while keeping the meaning of the text unchanged. 
\end{itemize}

\subsection{Demographics}
\begin{itemize}
    \item How frequently do you use any AI-powered chatbots?
%    \begin{itemize}
%        \item Every day
%        \item A few days per week
%        \item A few days per month
%        \item Less than a few days per month
%    \end{itemize}
    \item Did you know, before taking this survey, that an AI chatbot could possibly figure out personal information not explicitly shared in the text?
%    \begin{itemize}
%        \item Yes
%        \item No
%    \end{itemize}
    \item If yes: How did you come to know that a chatbot could figure out personal information not explicitly shared in text?
    \item What is your age?
%    \begin{itemize}
%        \item Under 18 years old
%        \item 18-24 years old
%        \item 25-34 years old
%        \item 35-44 years old
%        \item 45-54 years old
%        \item 55 years old or older
%    \end{itemize}
    \item What gender do you identify with?
%    \begin{itemize}
%        \item Male
%        \item Female
%        \item Non-binary / third gender
%        \item Prefer not to say
%        \item Prefer to self describe
%    \end{itemize}
    \item Which group(s) do you identify with?
    \item Do you live in the U.S.?
    \item What is your highest level of education attained?
    \item Which of the following industries most closely matches the one in which you are employed?
\end{itemize}

\section{Demographic Influence}\label{app:additional_analysis}
\begin{figure*}[ht]
\noindent
\centering
\begin{minipage}{.5\textwidth}
\centering
    \includegraphics[width=0.9\linewidth]{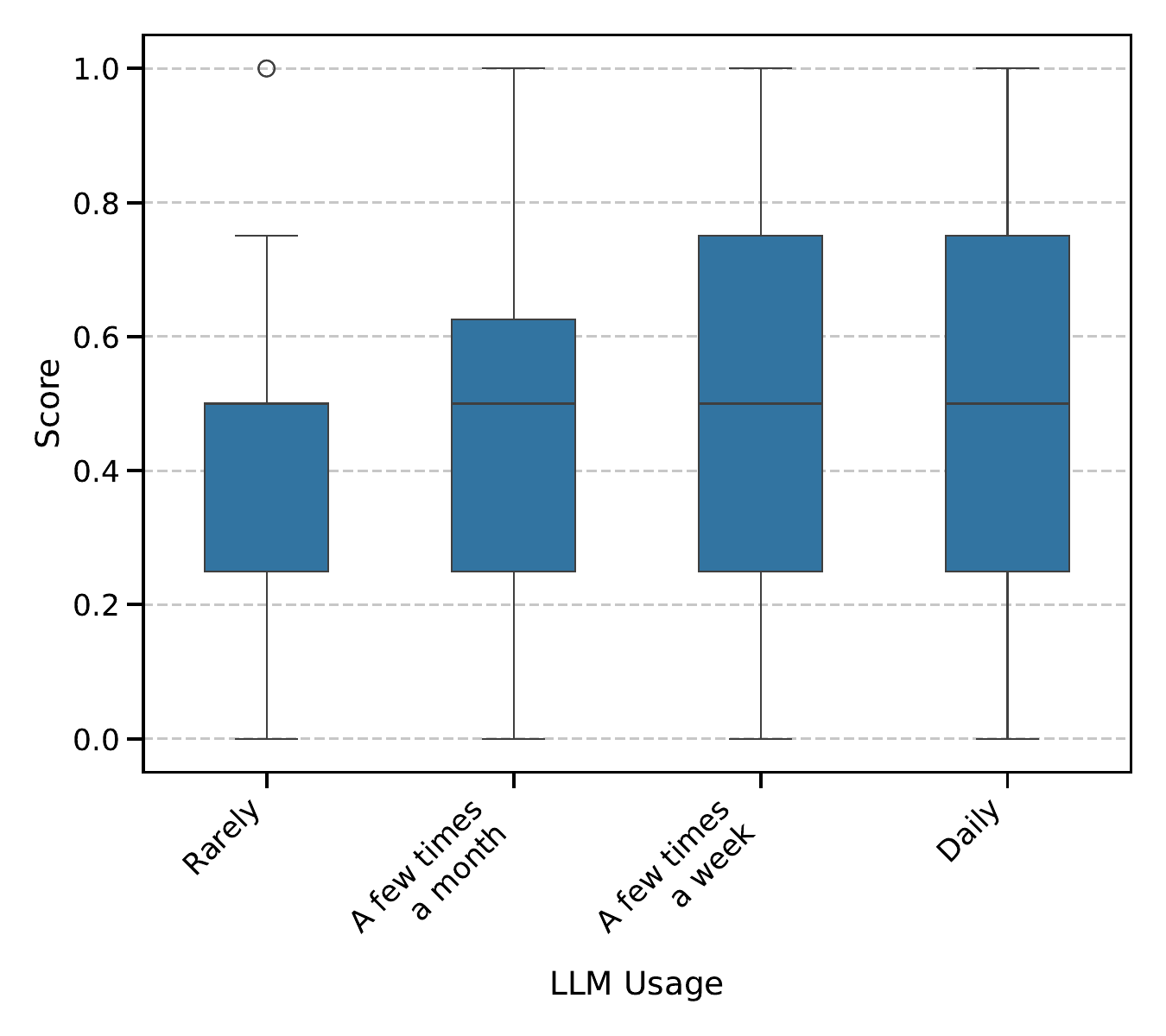}
    \vspace{-8pt}
    \caption{Distribution of score by usage frequency.}
    \label{fig:score_by_usage}
\end{minipage}\begin{minipage}{.5\textwidth}
\centering
    \includegraphics[width=0.9\linewidth]{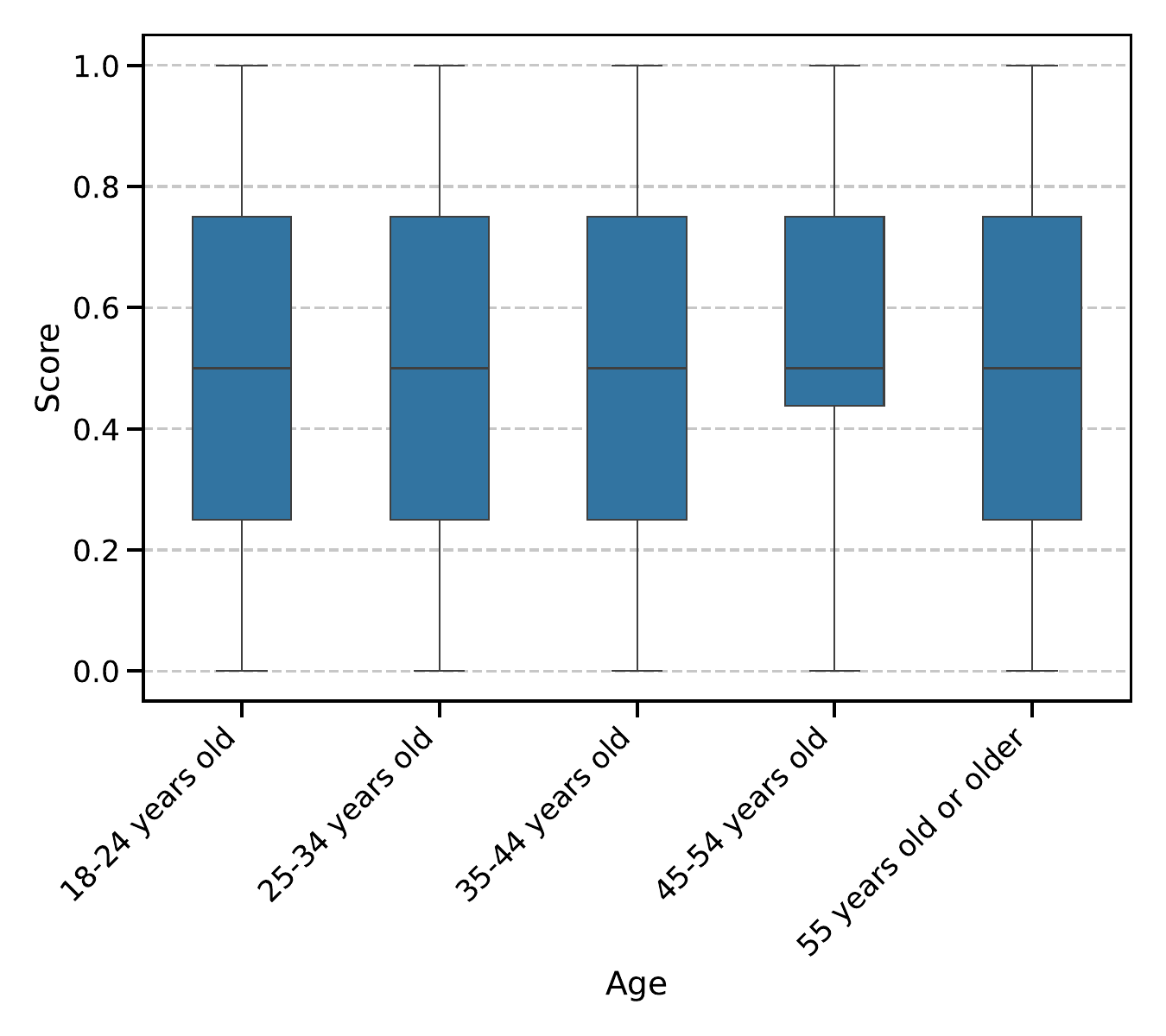}
    \vspace{-8pt}
    \caption{Distribution of score by age.}
    \label{fig:score_by_age}
\end{minipage}
\begin{minipage}{.5\textwidth}
\centering
    \includegraphics[width=0.9\linewidth]{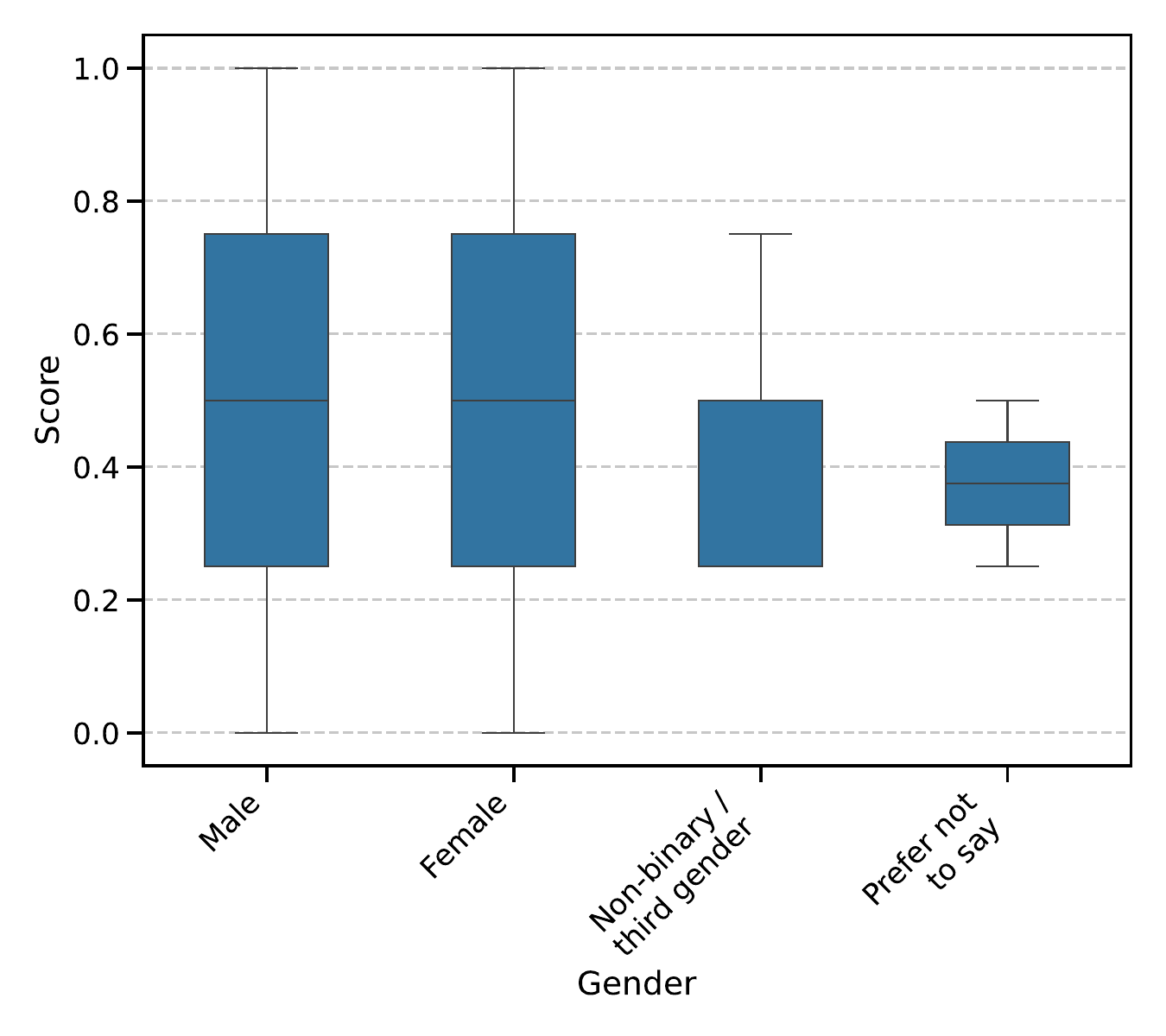}
    \vspace{-8pt}
    \caption{Distribution of score by gender.}
    \label{fig:score_by_gender}
\end{minipage}\begin{minipage}{.5\textwidth}
\centering
    \includegraphics[width=0.9\linewidth]{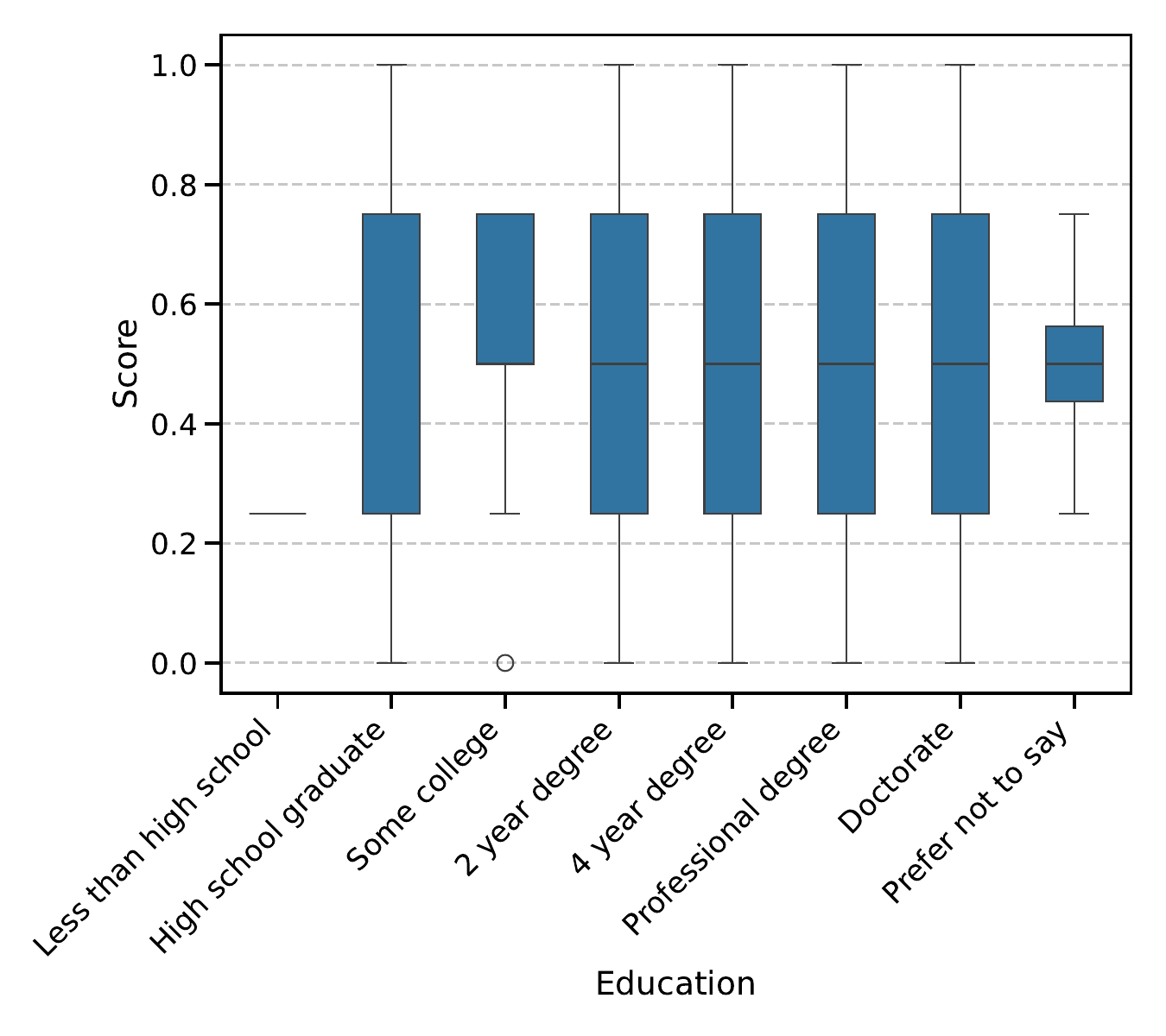}
    \vspace{-8pt}
    \caption{Distribution of score by education.}
    \label{fig:score_by_education}
\end{minipage}
\end{figure*}

We present additional plots on participant scores here. Figure~\ref{fig:score_by_usage},~\ref{fig:score_by_age},~\ref{fig:score_by_gender}, and~\ref{fig:score_by_education} shows the distribution of participant scores by how frequently participants used LLMs, their age, gender, and education level, respectively.

\end{document}